\documentclass[twoside,twocolumn,english,pra,show pacs]{revtex4}
\usepackage[T1]{fontenc}
\usepackage[latin9]{inputenc}
\usepackage{color}
\usepackage{amstext}
\usepackage{amssymb}
\usepackage{graphicx}
\usepackage{esint}

\makeatletter

\@ifundefined{textcolor}{}
{%
 \definecolor{BLACK}{gray}{0}
 \definecolor{WHITE}{gray}{1}
 \definecolor{RED}{rgb}{1,0,0}
 \definecolor{GREEN}{rgb}{0,1,0}
 \definecolor{BLUE}{rgb}{0,0,1}
 \definecolor{CYAN}{cmyk}{1,0,0,0}
 \definecolor{MAGENTA}{cmyk}{0,1,0,0}
 \definecolor{YELLOW}{cmyk}{0,0,1,0}
}

\usepackage{multirow}

\makeatother

\usepackage{babel}
\begin{document}

\title{Finite size effects and Hofstadter butterfly in a bosonic Mott insulator
\\with relativistic dispersion background}

\author{A. S. Sajna\textsuperscript{1}, T. P. Polak\textsuperscript{1}}
\begin{abstract}
Gauge potentials with different configurations have been recently
realized in the optical lattice experiments. It is remarkable that
one of the simplest gauge can generate particle energy spectrum with
the self-similar structure known as a Hofstadter butterfly. We investigate
theoretically the impact of strong on-site interaction on such a spectrum in the Bose-Hubbard model. In particular, it is shown that the fractal structure is encoded in
the quasi-particle and hole bosonic branches. A square lattice and
other structures (brick-wall and staggered magnetic flux lattice)
with relativistic energy dispersions which are currently accessible
in the experiments are considered. Moreover, although in brick-wall
and staggered flux lattices the quasi-particle densities of states
looks qualitatively similar, the corresponding Hofstadter butterfly
assumes different forms. In particular, we use a superposition of
two different synthetic gauge fields which appears to be a generator
of non-trivial phenomena in the optical lattice systems. The analysis
is carried out within the strong coupling expansion method on the
finite size lattices and also at finite temperatures which are relevant
for the currently made experiments.
\end{abstract}


\address{\textsuperscript{1}Faculty of Physics, Adam Mickiewicz University,
ul. Umultowska 85, 61-614 Pozna\'{n}, Poland}


\maketitle

\section{Introduction}

Ultracold atoms in an optical lattice have become very clean framework
for quantum simulations of condensed matter and high energy physics
\cite{RevModPhys.80.885,Lewenstein:2007vi}. In particular, optical
lattice systems offer the opportunity to study Abelian or non-Abelian
gauge fields. Some types of such potentials have been experimentally
realized in recent years \cite{2009Natur.462..628L,2011PhRvL.107y5301A,2013PhRvL.111r5301A,2013PhRvL.111r5302M,2012PhRvL.108v5304S,2013NatPh...9..738S,Kennedy2015,1610.06228,Stuhl2015,Mancini2015}.
An example of the non-trivial phenomena generated by the simplest
Abelian gauge (Landau configuration) is the Hofstadter butterfly \cite{1976PhRvB..14.2239H}.
It corresponds to the uniform magnetic field and implies fractal energy
spectrum. This spectrum has been investigated in a much broader context
of different gauges and lattice geometry modifications (see, e.g.
Refs \cite{2009PhRvA..79b3624G,Bermudez:2010id,2005PhRvL..95a0403O,1367-2630-5-1-356,PhysRevA.91.063628,1611.07052,2011PhRvA..83b3607Z,Agazzi2014,PhysRevA.89.011603}).
In order to get a more general theoretical description of the atoms
on an optical lattice in the systems with flux attached, the effects
caused by particles interactions should be taken into account \cite{Greiner:2002wt,Bakr:2010wn,Sherson:2010wz,2012Natur.487..454E}.
Therefore if we are interested in the bosonic dynamics in which the
single-particle spectrum is affected by the gauge fields \cite{2008PhRvL.100m0402L,2010PhRvA..81b3404L,2013PhRvL.111r5302M,2011PhRvL.107y5301A,2013PhRvL.111r5301A,Kennedy2015,2014PhRvA..89b3631S,PhysRevA.90.043603,Polak:2013jd},
we should take such correlations into account \cite{2011PhRvA..83a3612P,PhysRevLett.104.255303,2014PhRvA..89b3631S,PhysRevA.90.043603,PhysRevB.75.045133,PhysRevA.76.055601,2010PhRvA..82f3625M,Huang2011,2010PhRvA..81b3404L,2008PhRvL.100m0402L,PhysRevA.85.023622,1611.00205,2012PhRvA..85a3642S,Sachdeva:2010cd,PhysRevA.94.063628,PhysRevA.94.043606,PhysRevA.92.023609,PhysRevB.91.094502,PhysRevA.93.063610,PhysRevA.92.063630,Toga2015,PhysRevA.91.063611,PhysRevA.90.023608}.
It is interesting to note that recently the first experimental realization
of strongly interacting bosonic atoms in a gauge field was made on
the ladder lattice in the few body limit \cite{1612.05631}

Here we focus on the strongly correlated bosonic system described
by the Bose Hubbard model (BHM) \cite{1963PhRv..129..959G,Fisher:1989zza}.
We use the effective field theory to investigate the quasi-particle
spectrum of the bosonic Mott insulator (MI) phase. The analysis is
performed for the Landau gauge in the whole range of flux strength
per plaquette . So far only some chosen values of flux strength have
been investigated in the BHM within superfluid (SF) and MI spectra
\cite{2011PhRvA..83a3612P,PhysRevLett.104.255303,2014PhRvA..89b3631S,PhysRevA.90.043603}.
In particular, we analyze the appearance of the self-similar structure
in a collective behavior of strongly correlated bosons. 

Knowing that optical lattice patterns give us the opportunity to modify
the geometry of lattice structures, we show that such modifications
have a significant impact on the fractal-like structure in the MI
phase (in this work as an example we analyze square and brick-wall
lattices \cite{2012Natur.483..302T,PhysRevLett.110.165304}). Further,
we show that the tunability of self-similar spectrum of the strongly
interacting bosons can be widely extended and goes beyond the geometrical
type modifications of the lattice. Namely, one can consider a lattice
which is already equipped by the synthetic gauge field. To show this
we use one of the simplest gauge field configuration which is Abelian
and also has some free control parameter, i.e. staggered flux lattice.
Moreover, we deliberately choose the brick wall lattice and staggered
flux lattice for the analysis. Our aim is to show that although these
two types of lattices have qualitatively similar densities of states
(DOS) (with relativistic dispersion), they give very different structures
of the Hofstadter butterfly spectra. These differences are especially
enhanced when the free parameter of the staggered flux lattice is
tuned. 

It is important to stress, that our calculations include the finite
size effects which are present in the current experiments \cite{PhysRevLett.92.130403}.
Moreover, to properly establish the stability of MI phase in the parameter
space of BHM, the phase diagram analysis is done. We show how finite
size effects and the non-trivial lattices modify the critical line
which extends the previous works on this subject \cite{2010PhRvA..81b3404L,2008PhRvL.100m0402L,PhysRevB.75.045133,PhysRevA.85.023622}.

It is worth adding that very recently Hofstadter spectrum was investigated
in the hard-core limit of two component BHM \cite{1611.00205}. The
authors of this work have focused on the BHM topological properties.
In contrast, we investigate a finite interaction strength which enable
us to study the SF-MI phase boundary and the spectra of MI phase \cite{RevModPhys.80.885,Lewenstein:2007vi}.Besides,
we also consider the effects of thermal fluctuations on the bosonic
dynamics. Especially, we discuss their impact on the relativistic
part of the spectrum which can be useful for future experimental setups.

The paper is organized as follows, we first describe the model and
method applied (Sec. II). Next, in Sec. III, we use this method in
the analysis of the self-similarity of the quasi-particle excitations
for different lattices in bosonic MI phase. At the end of Sec. III
we also discuss the effects of thermal excitations in the experimentally
achievable range of the model parameters. Finally in Sec. IV we give
a summary of our work.

\section{The Method \label{sec:Model}}

\subsection{Effective action in MI phase for finite size lattice}

We consider BHM with the Hamiltonian given by

\begin{equation}
H=-\sum_{\langle ij\rangle}J_{ij}\hat{b}_{i}^{\dagger}\hat{b}_{j}+\frac{U}{2}\sum_{i}\hat{b}_{i}^{\dagger}\hat{b}_{i}^{\dagger}\hat{b}_{i}\hat{b}_{i}-\mu\sum_{i}\hat{b}_{i}^{\dagger}\hat{b}_{i}\,,\label{eq:BHM-hamiltonian}
\end{equation}
where $J_{ij}$, $U$, $\mu$ are the hopping energy, on-site interaction
energy and chemical potential, respectively. Besides, $\hat{b}_{i}$
and ($\hat{b}_{j}^{\dagger}$) is the annihilation and creation bosonic
operators at site i (the number of lattice sites is $N$). In the
coherent path integral representation, the partition function of BHM
takes the form 
\begin{equation}
\mathcal{Z}=\int\mathcal{D}b^{*}\mathcal{D}b\, e^{-\left(S_{0}+S_{1}\right)/\hbar},\label{eq:statistical-sum-0}
\end{equation}
\begin{eqnarray}
S_{0} & = & \sum_{i}\int_{_{0}}^{\hbar\beta}d\tau\left\{ b_{i}^{*}(\tau)\hbar\partial_{\tau}b_{i}(\tau)\right.\nonumber \\
 &  & \left.+\frac{U}{2}b_{i}^{*}(\tau)b_{i}^{*}(\tau)b_{i}(\tau)b_{i}(\tau)-\mu b_{i}^{*}(\tau)b_{i}(\tau)\right\} .
\end{eqnarray}
\begin{equation}
S_{1}=-\sum_{\langle ij\rangle}\int_{_{0}}^{\hbar\beta}d\tau\; J_{ij}b_{i}^{*}(\tau)b_{j}(\tau).\label{eq:actionprzed3}
\end{equation}
where $b_{i}(\tau)$ is the complex field over imaginary time and
$\beta$ is the inverse of temperature $1/k_{B}T$ ($k_{B}$ is the
Boltzmann constant). To obtain the effective quadratic action in the
Mott insulator phase, we employ the strong coupling method from \cite{2005PhRvA..71c3629S}.
This method assumes perturbative treatment of $S_{1}$ in which the
double Hubbard-Stratonovich transformation together with cumulant
expansion are used. Then the second order effective action in $b$,
$\bar{b}$ fields reads
\begin{eqnarray}
S^{eff} & = & -\sum_{ij}\int d\tau dd\tau'J_{ij}b_{i}^{*}(\tau)b_{j}(\tau)\nonumber \\
 &  & -\hbar G_{0}^{-1}\left(\tau-\tau'\right)\sum_{i}\int d\tau d\tau'b_{i}^{*}(\tau)b_{i}(\tau')\label{eq:Seff1}
\end{eqnarray}
where 
\begin{equation}
G_{0}^{-1}\left(\tau-\tau'\right)=-\left\langle b_{i}^{*}(\tau)b_{i}(\tau')\right\rangle _{0}\label{eq:local-green}
\end{equation}
 and $\left\langle ...\right\rangle _{0}=Z_{0}^{-1}\int\mathcal{D}b^{*}\mathcal{D}b\,...\, e^{-S_{0}/\hbar}$
with $Z_{0}=\int\mathcal{D}b^{*}\mathcal{D}b\, e^{-S_{0}/\hbar}$
(see also Appendix \ref{sub:Local-Green-function}). In the Matsubara
frequency representation $\omega_{m}=2\pi\beta/m$ ($m\in\mathbb{Z}$),
Eq. (\ref{eq:Seff1}) yields diagonal form 
\begin{equation}
S^{eff}=-\sum_{ij}\sum_{m}\bar{b}_{im}\left[J_{ij}+\hbar G_{0}^{-1}\left(i\omega_{m}\right)\delta_{ij}\right]b_{jm}.\label{eq:Seff2}
\end{equation}
which can be rewritten in the matrix representation as follows 
\begin{equation}
S^{eff}=-\sum_{m}B_{m}^{\dagger}\left[\mathbf{J}+\hbar G_{0}^{-1}\left(i\omega_{n}\right)\mathbf{I}\right]B_{m}\label{eq:Seff3}
\end{equation}
where we denote $B_{m}=\left[b_{1m},b_{2m},...,b_{Nm}\right]^{T}$.
Now, it is explicitly seen, that the problem of evaluating the effective
action from Eq. (\ref{eq:Seff3}) reduces only to dealing with the
free particle non-diagonal part of the action $S^{eff}$ i.e. $\mathbf{J}$,
because $G_{0}^{-1}\left(i\omega_{n}\right)\mathbf{I}$ is already
a diagonal matrix. This is a key point of our calculations and it
will be discussed more explicitly later on.

For a hopping matrix $\mathbf{J}$ in Eq. (\ref{eq:Seff3}), one can
perform unitary transformation $B_{m}=U^{\dagger}\Phi_{m}$
\begin{equation}
S^{eff}=-\sum_{m}\Phi_{m}^{\dagger}\left[\mathbf{J}_{d}+\hbar G_{0}^{-1}\left(i\omega_{m}\right)\mathbf{I}\right]\Phi_{m}\label{eq:Seff4}
\end{equation}
with $\mathbf{J}_{d}=U\mathbf{J}U^{\dagger}$ where $\mathbf{J}_{d}$
is the diagonal matrix with eigenvalues $\varepsilon_{\lambda}$ ,
$\lambda\in\left\{ 1,2,...,N\right\} $ and we denote $\Phi_{n}=\left[\phi_{1m},\phi_{2m},...,\phi_{Nm}\right]^{T}$.

\subsection{Finite size phase diagram at zero temperature}

The phase diagram boundary is calculated from the vanishing of the
second order coefficient in the effective action from Eq. (\ref{eq:Seff4}),
i.e.
\begin{equation}
\varepsilon_{\lambda,\textrm{min}}+\hbar G_{0}^{-1}\left(i\omega_{m}=0\right)\mathbf{I}=0\label{eq:phase-boundary}
\end{equation}
where this condition is met at the static limit $i\omega_{m}=0$ and
for the lowest eigenvalues of $\mathbf{J}_{d}$ (which we denote by
$\varepsilon_{\lambda,\textrm{min}}$). The phase boundary in terms
of Eq. \ref{eq:phase-boundary} has a mean-field character therefore
we only consider the zero temperature limit in our two dimensional
considerations \cite{PhysRev.158.383}. Then, it is enough to use
three state approximation for the local Green function $G_{0}\left(i\omega_{m}\right)$
whose details are given in Appendix \ref{sub:Local-Green-function}.

\subsection{Quasi-particle density of states at zero temperature \label{sub:Single-particle-density T=00003D0}}

If we are interested in the quasi-particle DOS in the MI phase, we
firstly define Green's function in the Mott insulator regime
\begin{equation}
\mathcal{G}^{MI}\left(\omega_{m}\right)=-\left\langle B_{m}^{\dagger}B_{m}\right\rangle =-\left\langle \Phi_{m}^{\dagger}\Phi_{m}\right\rangle =\sum_{_{\lambda}}G^{MI}\left(\omega_{m},\varepsilon_{\lambda}\right).
\end{equation}
where 
\begin{eqnarray}
G^{MI}\left(\omega_{m},\varepsilon_{\lambda}\right) & = & -\left\langle \phi_{\lambda m}^{*}\phi_{\lambda m}\right\rangle =\frac{G_{0}^{TSA}(i\omega_{m})}{1+\varepsilon_{\lambda}G_{0}^{TSA}(i\omega_{m})}\nonumber \\
 & = & \frac{z\left(\varepsilon_{\lambda}\right)}{i\omega_{m}-E^{+}(\varepsilon_{\lambda})}+\frac{1-z\left(\varepsilon_{\lambda}\right)}{i\omega_{m}-E^{-}(\varepsilon_{\lambda})}\,,\label{eq:GMI-start}
\end{eqnarray}
with
\begin{eqnarray}
E^{\pm}\left(\varepsilon_{\lambda}\right) & = & -\frac{\varepsilon_{\lambda}}{2}-\mu+U\left(n_{0}+\frac{1}{2}\right)\nonumber \\
 &  & \pm\frac{1}{2}\sqrt{\text{\ensuremath{\varepsilon_{\lambda}}}^{2}-4\varepsilon_{\lambda}U\left(n_{0}+\frac{1}{2}\right)+U^{2}},
\end{eqnarray}
\begin{equation}
z\left(\varepsilon_{\lambda}\right)=\frac{E^{+}\left(\varepsilon_{\lambda}\right)+\mu+U}{E^{+}\left(\varepsilon_{\lambda}\right)-E^{-}\left(\varepsilon_{\lambda}\right)}.\label{eq:GMI-end}
\end{equation}
and $G_{0}\left(i\omega_{m}\right)$ is given in Eq. (\ref{eq:TSAg0}).
Then the DOS is calculated within the standard procedure 
\begin{eqnarray}
\rho_{MI}^{TSA}\left(\omega\right) & = & -\frac{1}{\pi N}\sum_{\lambda=1}^{N}\textrm{Im}\left[\frac{z\left(\varepsilon_{\lambda}\right)}{\omega-E^{+}(\varepsilon_{\lambda})+i\eta}\right.\nonumber \\
 &  & \left.+\frac{1-z\left(\varepsilon_{\lambda}\right)}{\omega-E^{-}(\varepsilon_{\lambda})+i\eta}\right]\,,\label{eq: main-result}
\end{eqnarray}
where $\eta$ is the spectrum broadening parameter. It is important
to stress that the form of Eqs. (\ref{eq:GMI-start}-\ref{eq:GMI-end})
corresponds to the standard form known in the Random Phase Approximation
(RPA) method for the MI Green function in which eigenenergies of $\mathbf{J}_{d}$
in Eq. (\ref{eq:Seff4}) are numerated by the wave vector $\mathbf{k}$
\cite{2005PhRvA..71c3629S,Freericks-strong-coupling,Sajna2015}.

Moreover, from the equation
\begin{eqnarray}
\rho\left(\omega\right) & = & -\frac{1}{\pi N}\sum_{\lambda=1}^{N}\textrm{Im}\frac{1}{\omega-\varepsilon_{\lambda}+i\eta}\label{eq: FREE DOS}
\end{eqnarray}
we calculate the free single-particle DOS. \cite{2006PhRvB..74l5116C} 

\begin{figure}[th]
\includegraphics[scale=0.33]{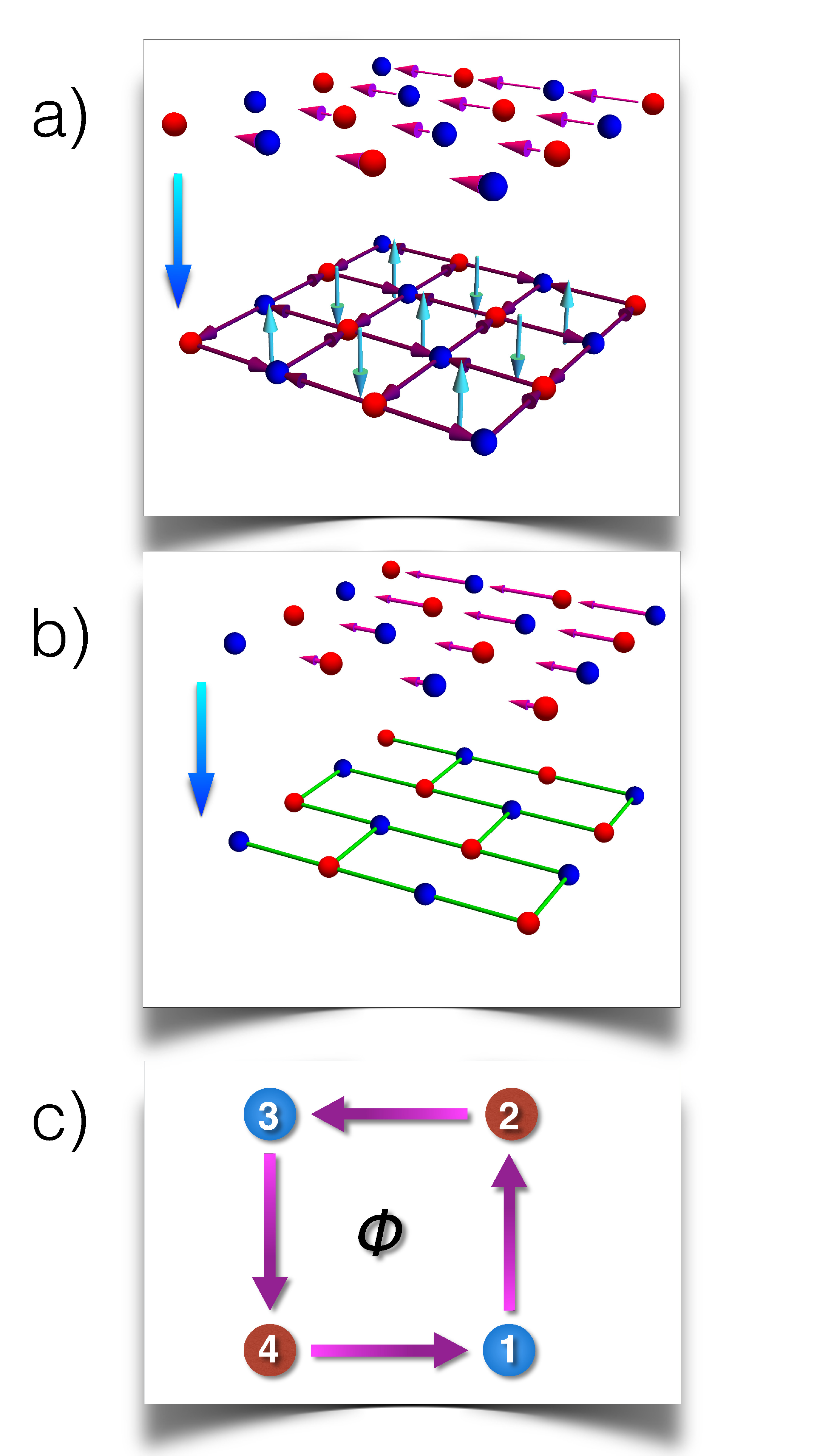}

\caption{(color online) Schematic representation of lattices and gauges investigated
in this work. (a) Staggered-flux lattice (below) with imposed Landau
gauge (above), (b) Brick-wall lattice (below) with imposed Landau
gauge (above), (c) Single plaquette with staggered flux gauge $\mbox{\ensuremath{\mathbf{A}}}_{S}$
Lattices and Landau gauge are plotted separately for clarity. Light
(dark) violet arrows between neighbour sites represent Landau $\mathbf{A}_{L}$
(staggered flux $\mbox{\ensuremath{\mathbf{A}}}_{S}$) gauge.}
\label{fig: gauges}
\end{figure}

\begin{figure}[th]
\includegraphics[scale=0.65]{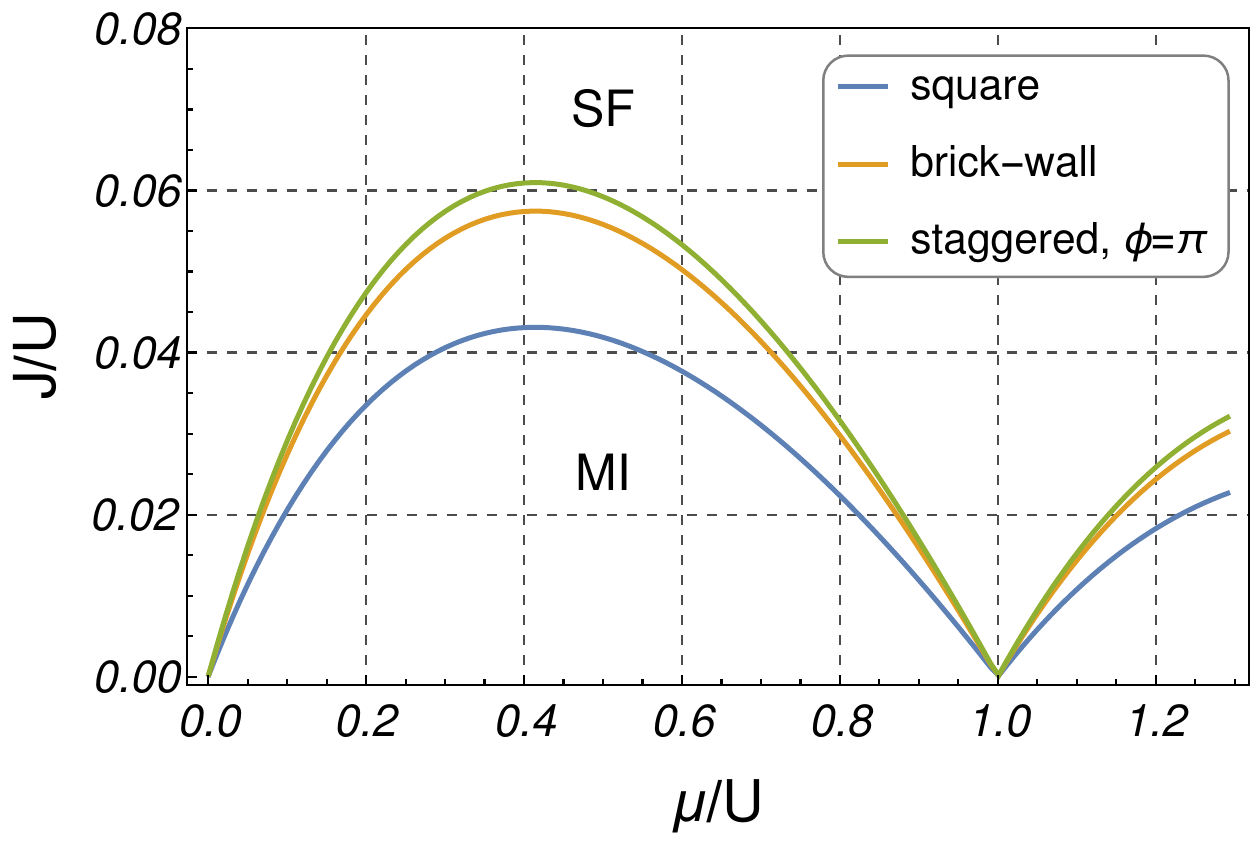}
\caption{(color online) Phase diagram of the Bose-Hubbard model for three different
type of lattices: square lattice (green line), brick-wall (orange
line) and lattice with staggered magnetic field (blue line). All lattices
have the finite size 30x30.}
\label{fig: pb-lobes}
\end{figure}

\begin{figure*}[th]
\includegraphics[scale=0.7]{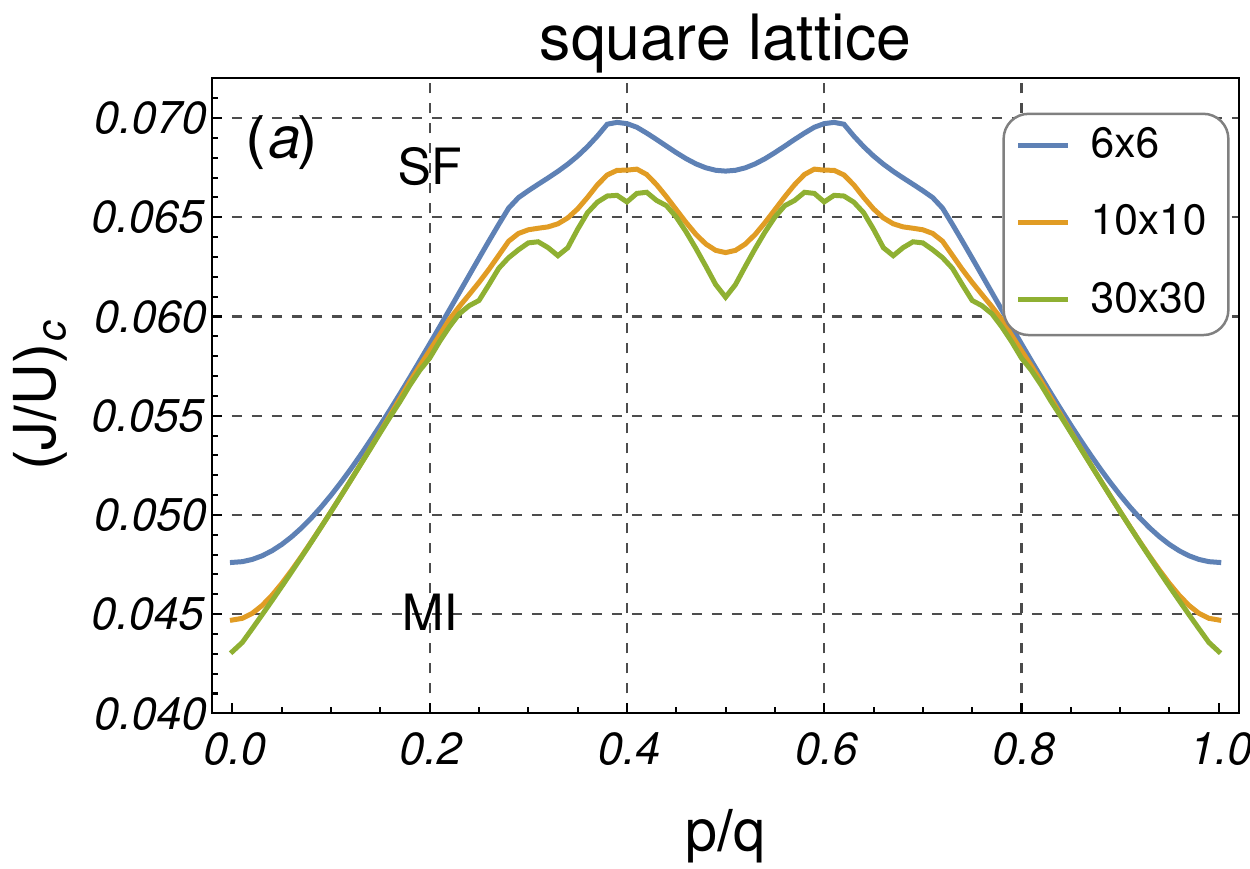}
\includegraphics[scale=0.7]{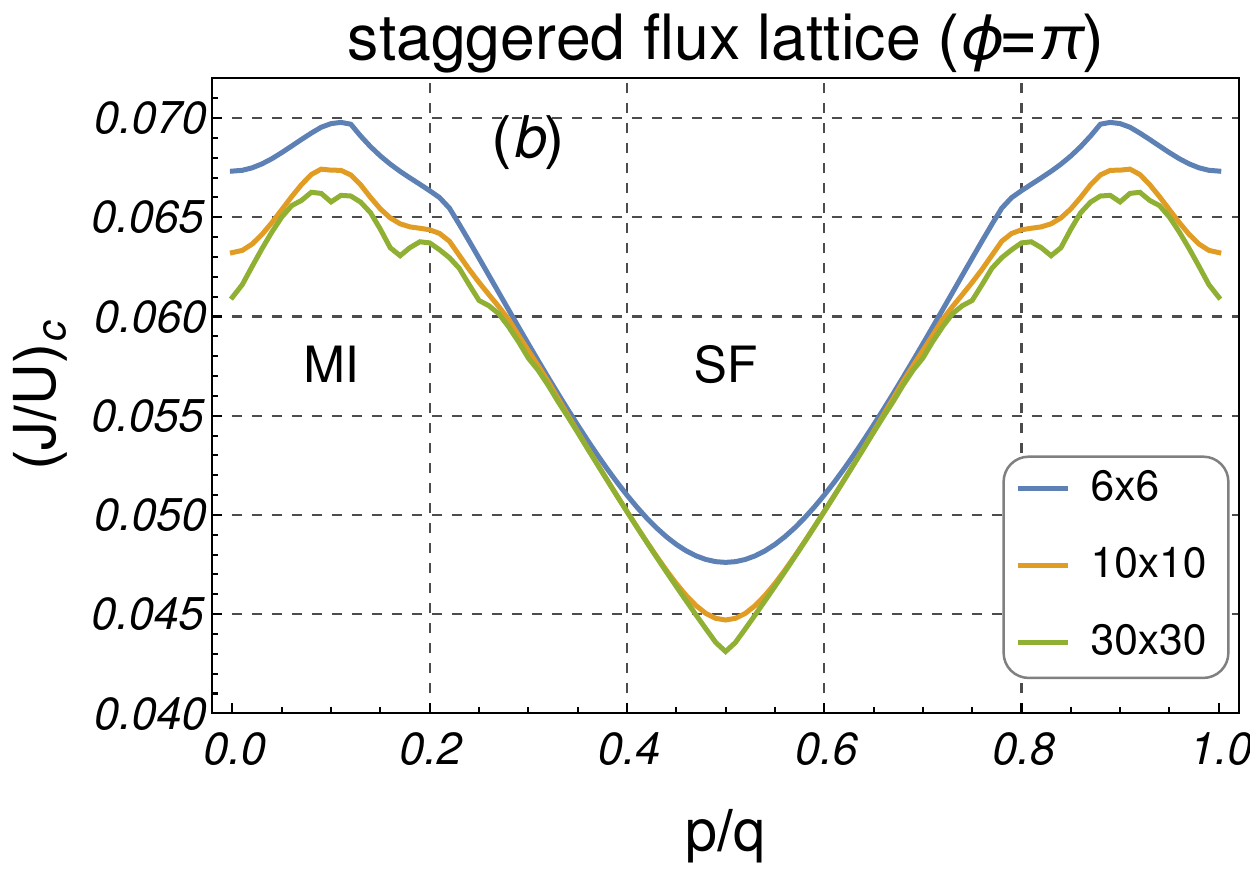}

\includegraphics[scale=0.7]{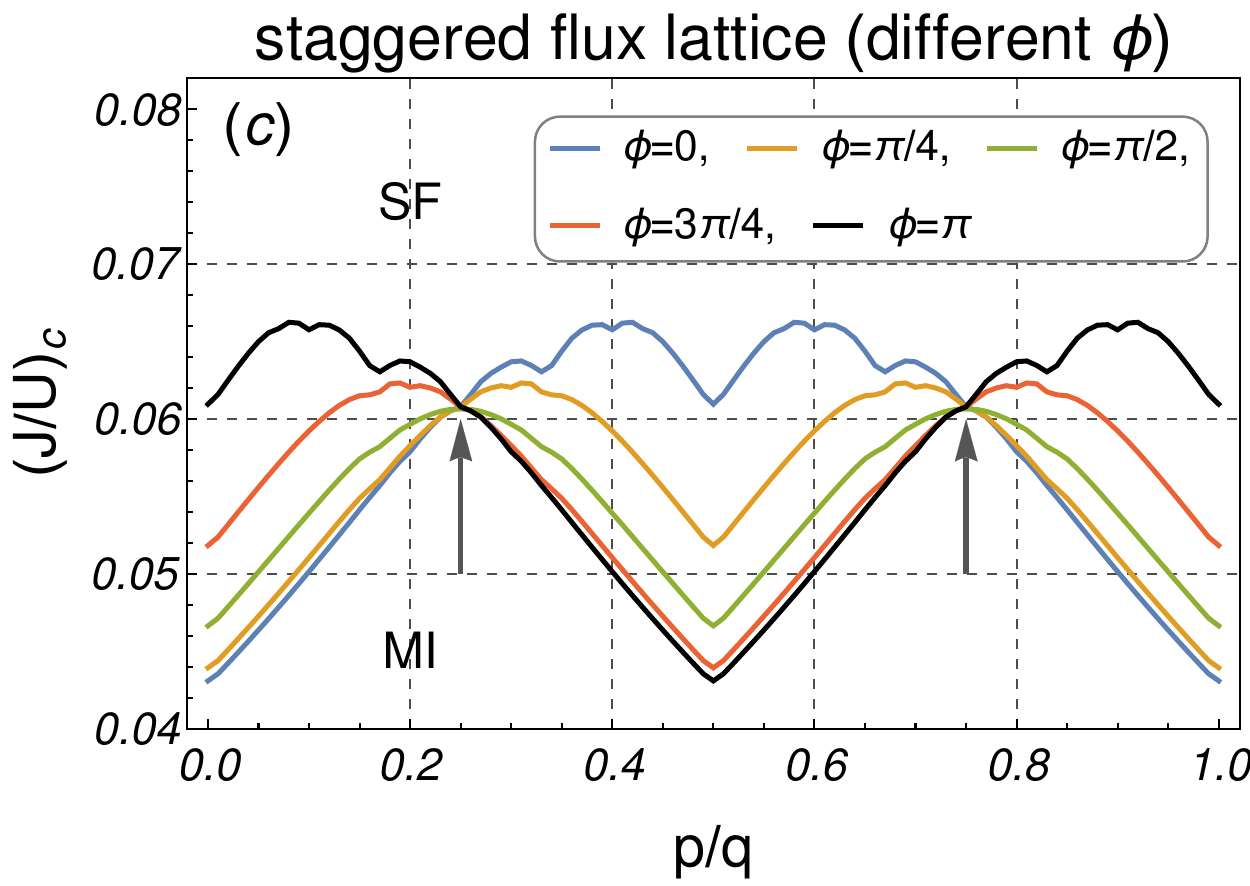}
\includegraphics[scale=0.7]{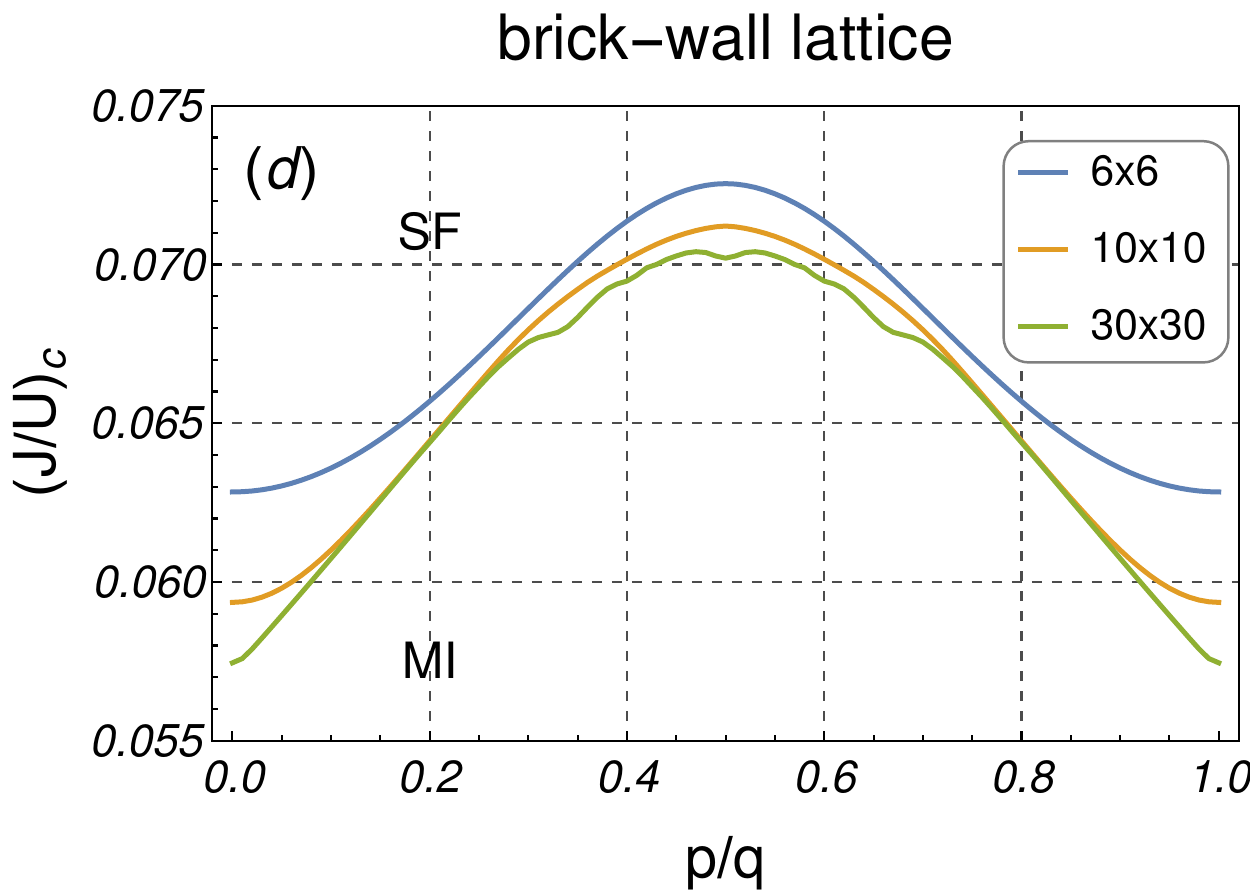}

\caption{(color online) Critical value of the hopping energy $(J/U)_{c}$ versus
flux per plaquette $p/q$. In Figs. (a-c), the phase boundary is plotted
for different amplitude of staggered flux $\phi$: (a) $\phi=0$ which
corresponds to standard square lattice, (b) $\phi=\pi$ which corresponds
to the highest value of magnetic flux per plaquette, (c) different
value of $\phi$ are plotted for comparison ($\phi=0$ - blue line,
$\phi=\pi/4$ - orange line, $\phi=\pi/2$ - green line, $\phi=3\pi/4$
- red line, $\phi=\pi$ - black line). In Fig. (d) phase boundary
for the brick-wall lattice is plotted. Moreover in Figs. (a), (b)
and (d) different lattice size are depicted, 6x6 (blue line), 10x10
(orange line), 30x30 (green line). All plots are made at the tip of
the first lobe in which $\mu/U=\sqrt{2}-1$. \label{fig:phase boundary at tip}}
\end{figure*}

\begin{figure}[th]
\includegraphics[scale=0.65]{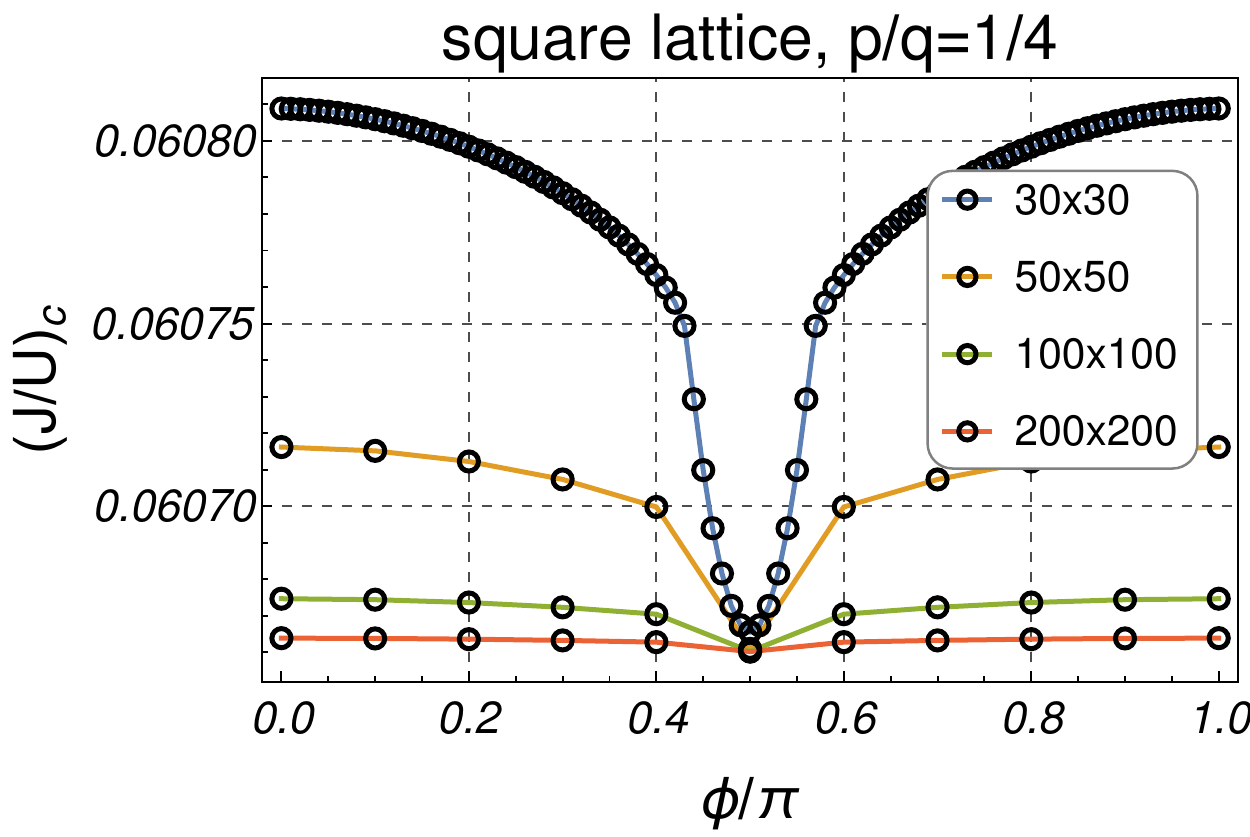}
\caption{(color online) Critical value of the hopping energy $(J/U)_{c}$ versus
$\phi/\pi$ for the lattice with additional uniform magnetic field
at $p/q=1/4$. The plot is drawn at the tip of the first lobe in which
$\mu/U=\sqrt{2}-1$.}
\label{fig: pb-special-point-1/4}
\end{figure}

\begin{figure}[th]
\includegraphics[scale=0.24]{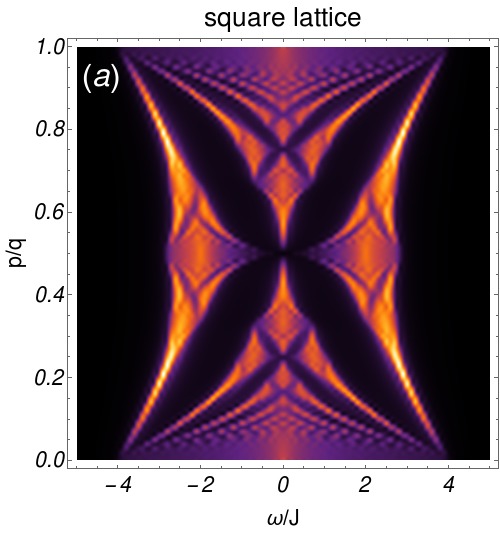}
\includegraphics[scale=0.24]{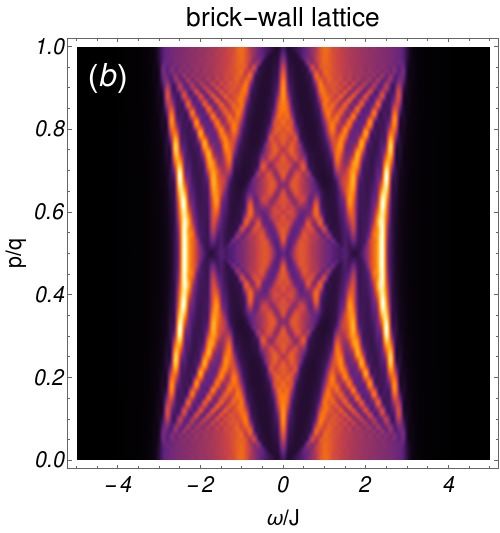}
\caption{(color online) Free particle density of states in $p/q-\omega/J$
plane (lattice size: $30\times30$). (a) and (b) are Hofstadter butterflies
for non-interacting bosons on the square and brick-wall lattice, respectively.
We set $\eta=0.06J$.}
\label{fig: butterfly FREE}
\end{figure}

\begin{figure*}[th]
\includegraphics[scale=0.24]{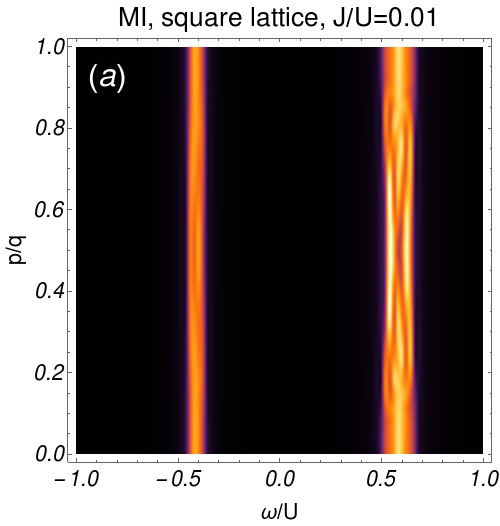}
\includegraphics[scale=0.24]{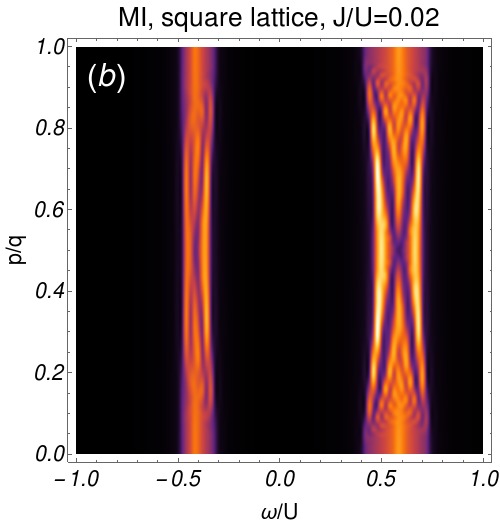}
\includegraphics[scale=0.24]{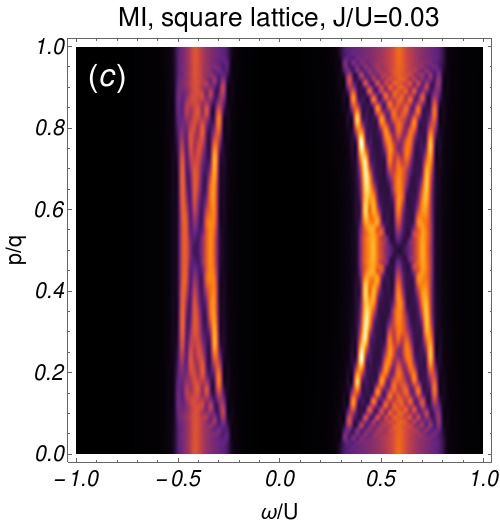}
\includegraphics[scale=0.24]{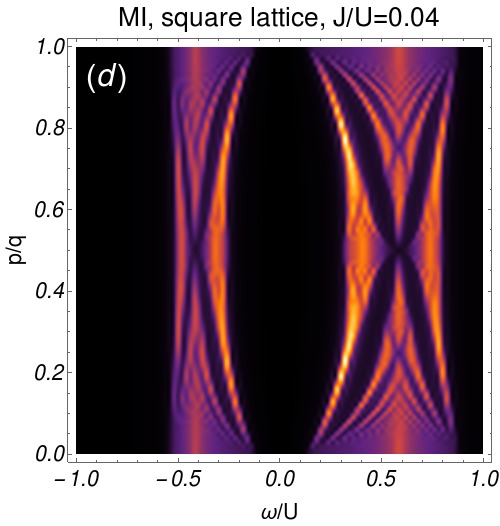}

\includegraphics[scale=0.24]{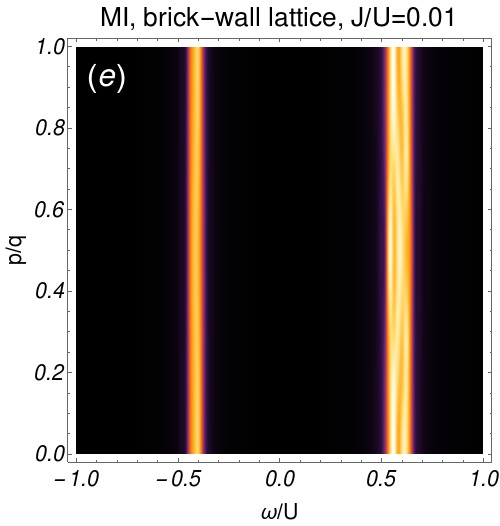}
\includegraphics[scale=0.24]{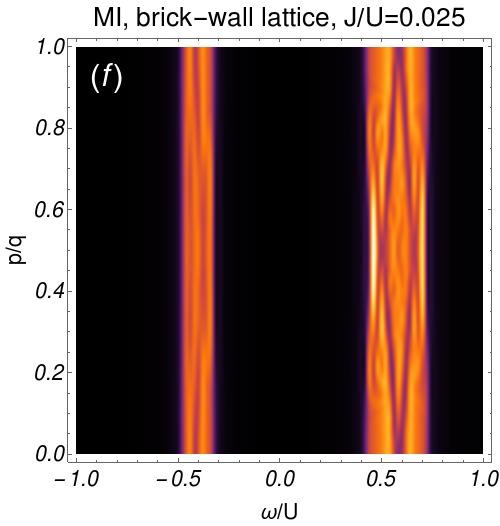}
\includegraphics[scale=0.24]{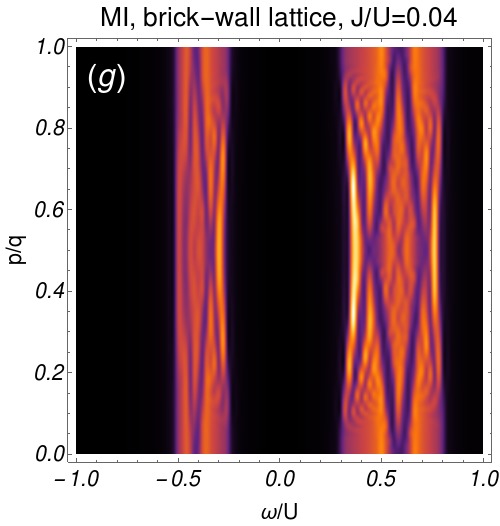}
\includegraphics[scale=0.24]{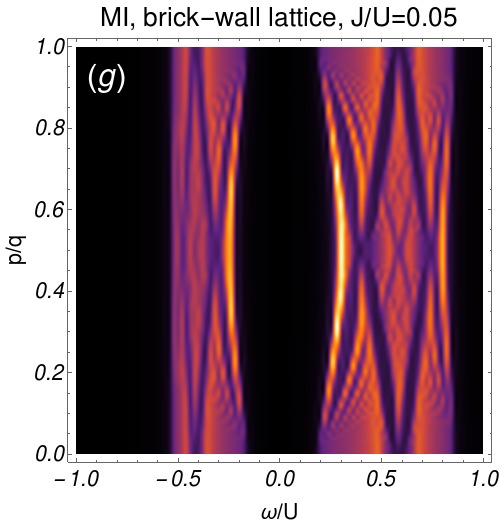}

\caption{(color online) Absolute value of density of states in the $p/q-\omega/U$
plane. (a-d) and (e-h) are Hofstadter butterflies in the Mott insulator
phase on the square and brick-wall lattice, respectively. The calculations
are performed for $30\times30$ lattice sites. BHM Hamiltonian parameters
are $\mu/U\approx0.41$ and (a) $J/U=0.01$, (b) $J/U=0.02$, (c)
$J/U=0.03$, (d) $J/U=0.04$, (e) $J/U=0.01$, (f) $J/U=0.025$, (g)
$J/U=0.04$, (h) $J/U=0.052$. Moreover, we set $\eta=0.01U$.\label{fig:butterfly}}
\end{figure*}

\begin{figure*}[th]
\includegraphics[scale=0.24]{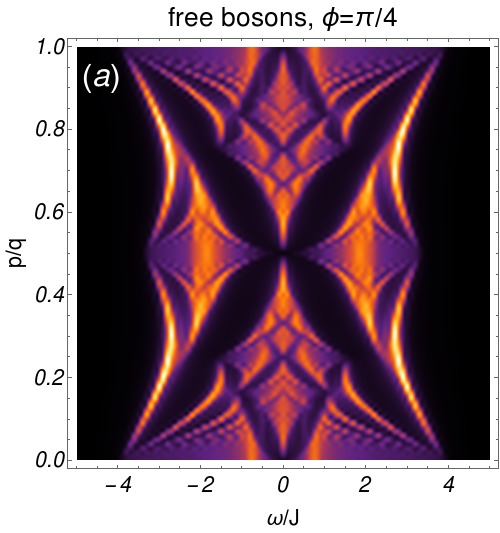}
\includegraphics[scale=0.24]{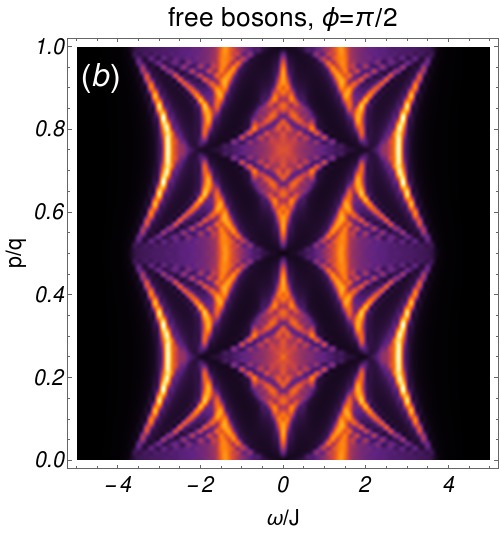}
\includegraphics[scale=0.24]{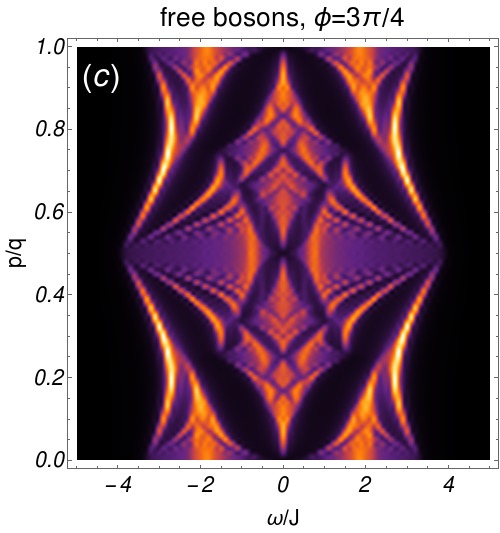}
\includegraphics[scale=0.24]{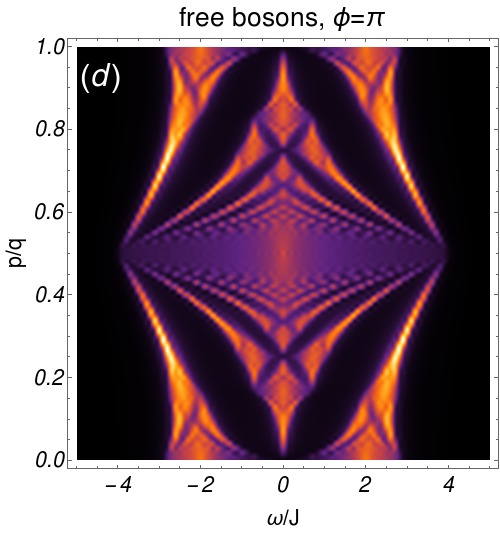}

\includegraphics[scale=0.24]{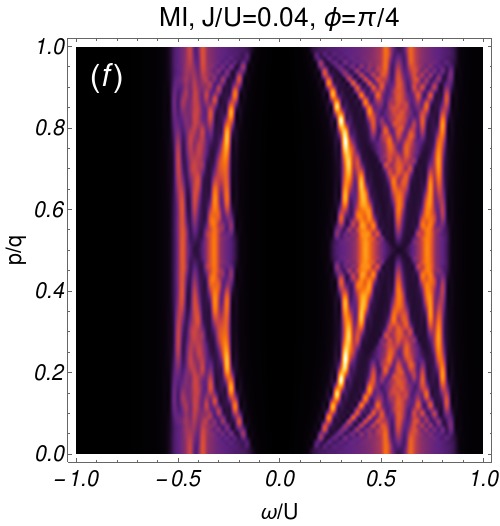}
\includegraphics[scale=0.24]{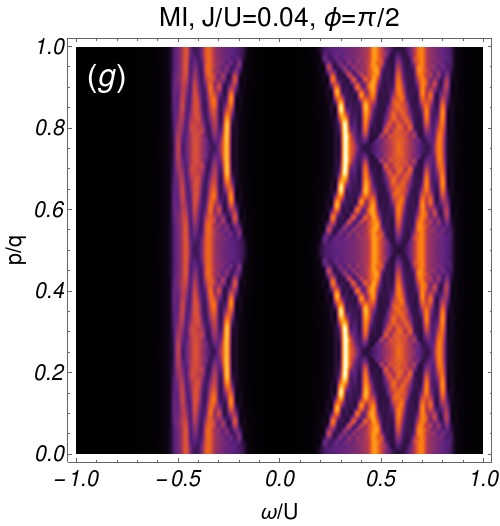}
\includegraphics[scale=0.24]{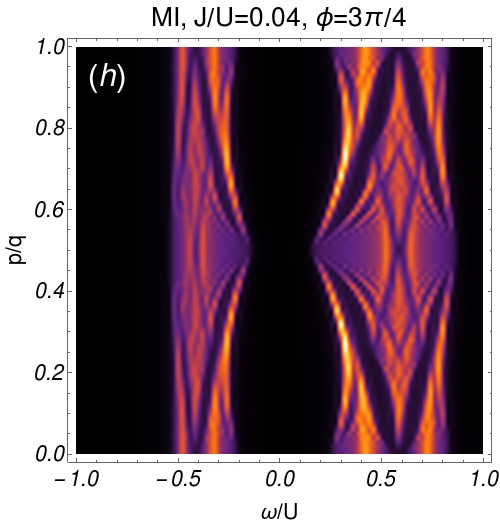}
\includegraphics[scale=0.24]{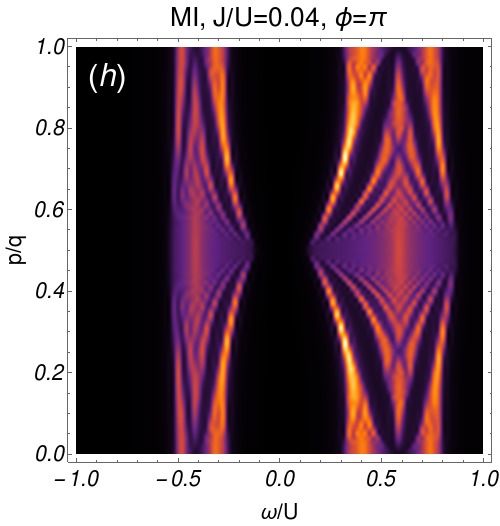}

\caption{(color online) Absolute value of density of states in the $p/q-\omega/U$
plane. (a-d) and (e-h) are Hofstadter butterflies in the Mott insulator
phase on the square and brick-wall lattice, respectively. The calculations
are made for $30\times30$ lattice sites. BHM Hamiltonian parameters
are $\mu/U\approx0.41$ and (a) $J/U=0.01$, (b) $J/U=0.02$, (c)
$J/U=0.03$, (d) $J/U=0.04$, (e) $J/U=0.01$, (f) $J/U=0.025$, (g)
$J/U=0.04$, (h) $J/U=0.052$. Moreover, we set $\eta=0.01U$. Moreover,
we set $\eta=0.06U$.\label{fig:butterfly-1}}
\end{figure*}

\begin{figure}[th]
\includegraphics[scale=0.7]{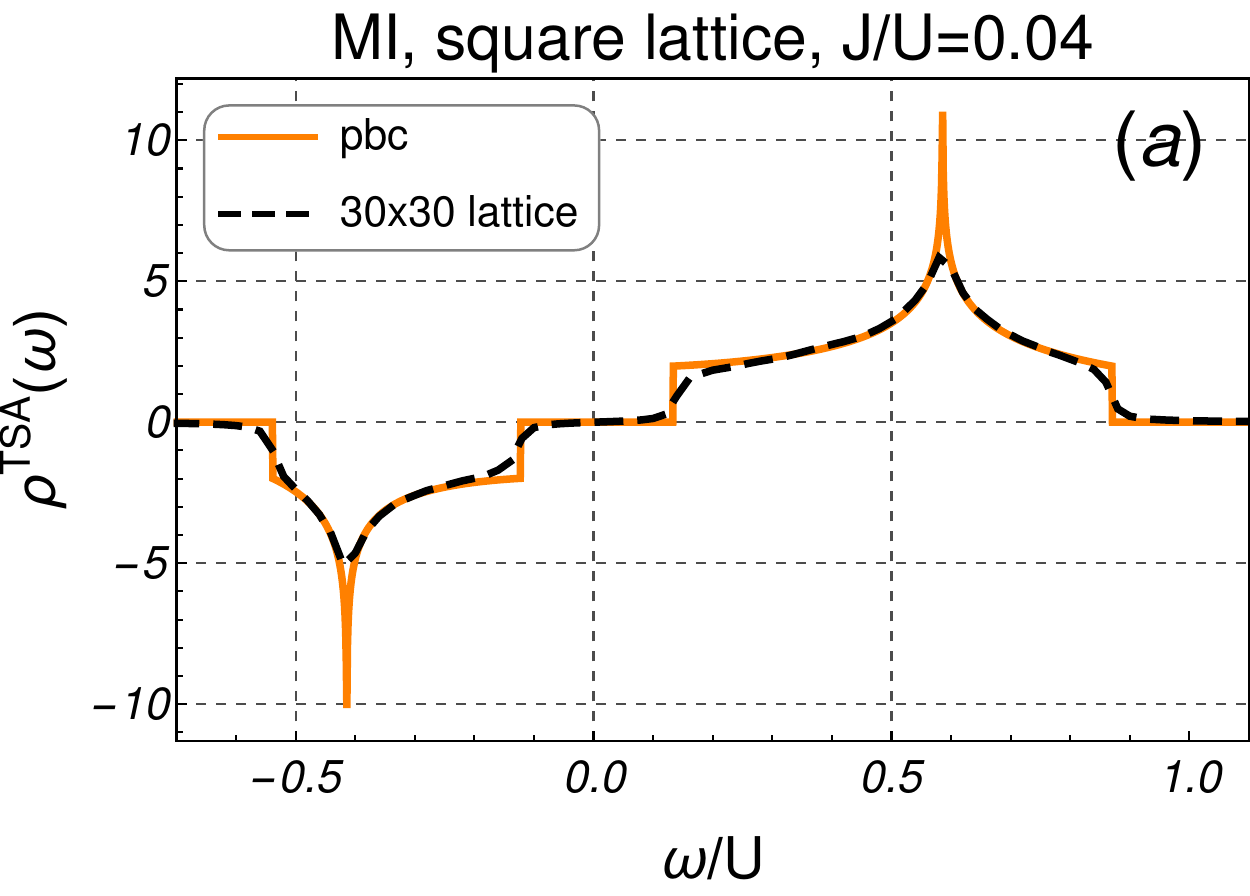}

\includegraphics[scale=0.7]{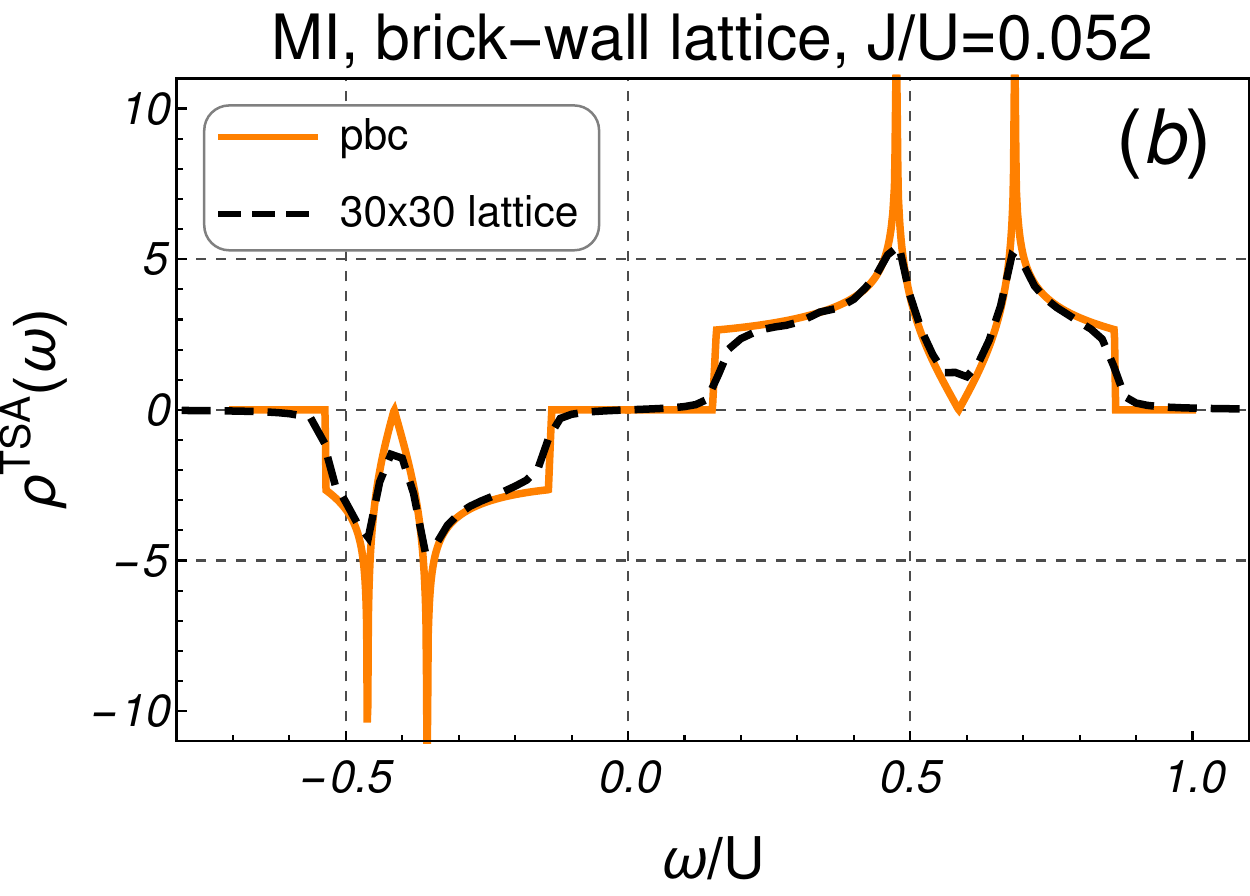}

\caption{(color online) Density of states in the MI phase for the square (a)
and brick-wall (b) lattices. Both plots are drawn for $p/q=0$. The
solid orange line corresponds to the exact density of states for the
lattice with periodic boundary conditions (pbc) (see, Appendix \ref{sub:Exact-density-of})
and the dashed black line corresponds to the finite size lattice $30\times30$.
Moreover, we set $\eta=0.01$ and (a) $J/U=0.04$, (b) $J/U=0.052$.
\label{fig: DOS exact - square and brick-wall}}
\end{figure}

\begin{figure}[th]
\includegraphics[scale=0.7]{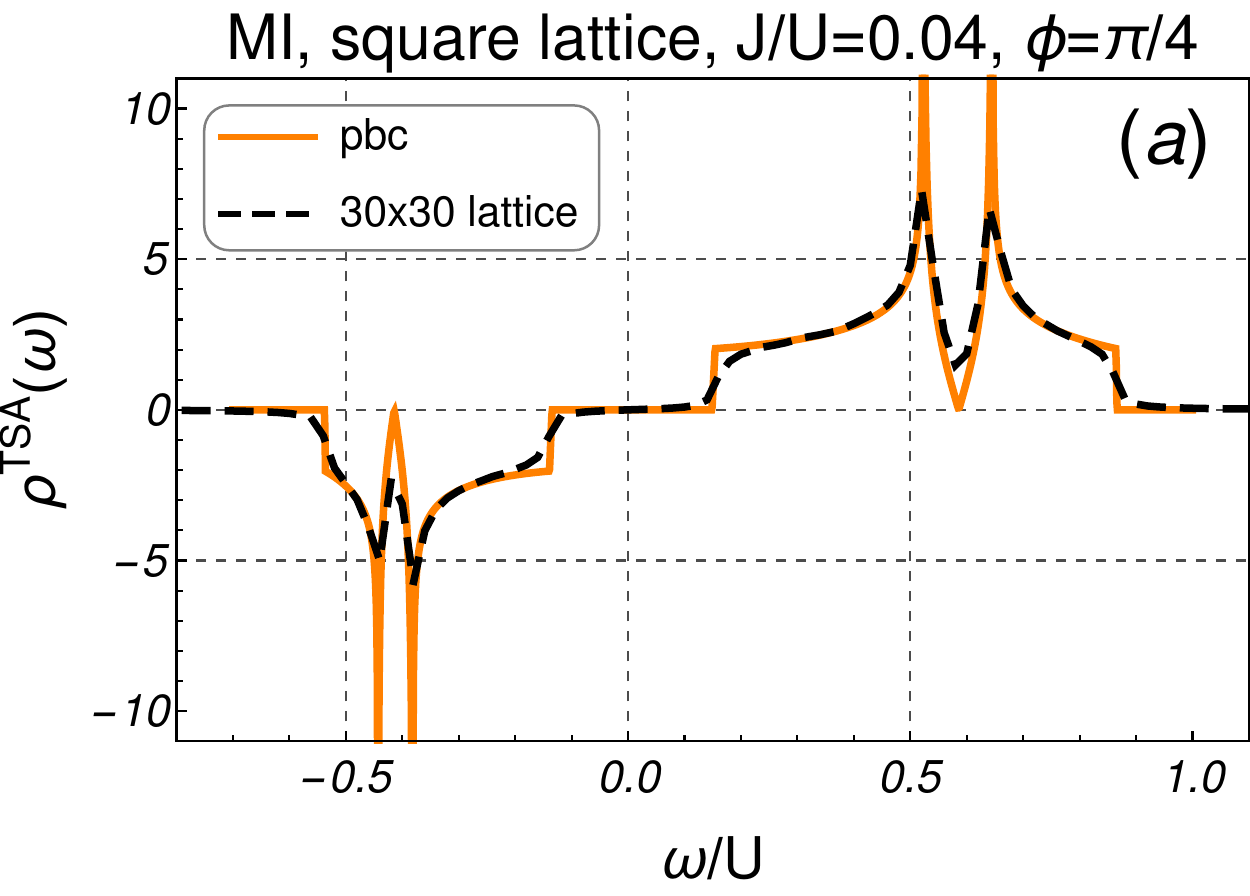}

\includegraphics[scale=0.7]{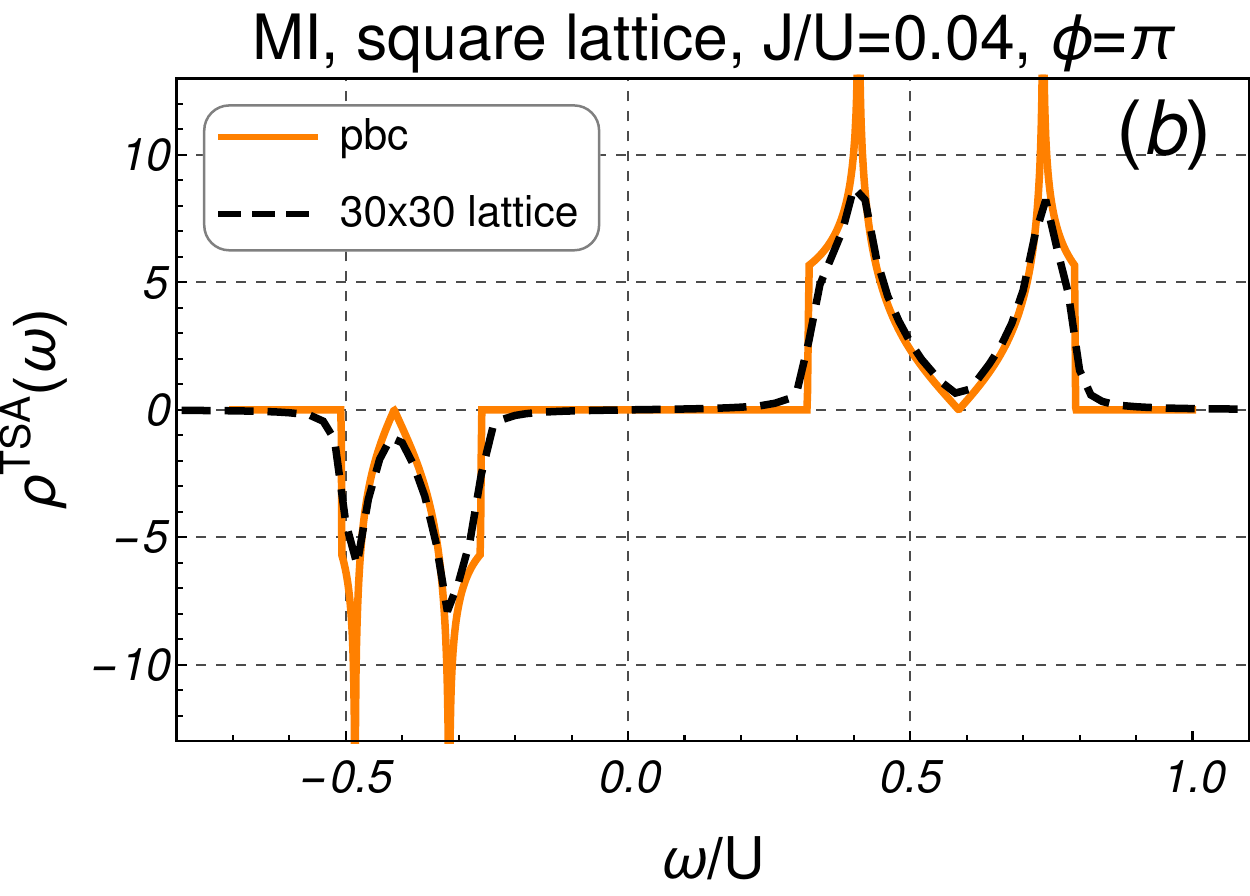}

\caption{(color online) Density of states in the MI phase for the square (a)
and brick-wall (b) lattices. Both plots are done for $p/q=0$. The
solid orange line correspond to exact density of states for the lattice
with periodic boundary conditions (pbc) (see, Appendix \ref{sub:Exact-density-of})
and dashed black line corresponds to the finite size lattice $30\times30$.
Moreover, we set $\eta=0.01$ and (a) $J/U=0.04$, (b) $J/U=0.052$.
\label{fig: DOS exact - staggered}}
\end{figure}

\begin{figure*}[th]
\includegraphics[scale=0.45]{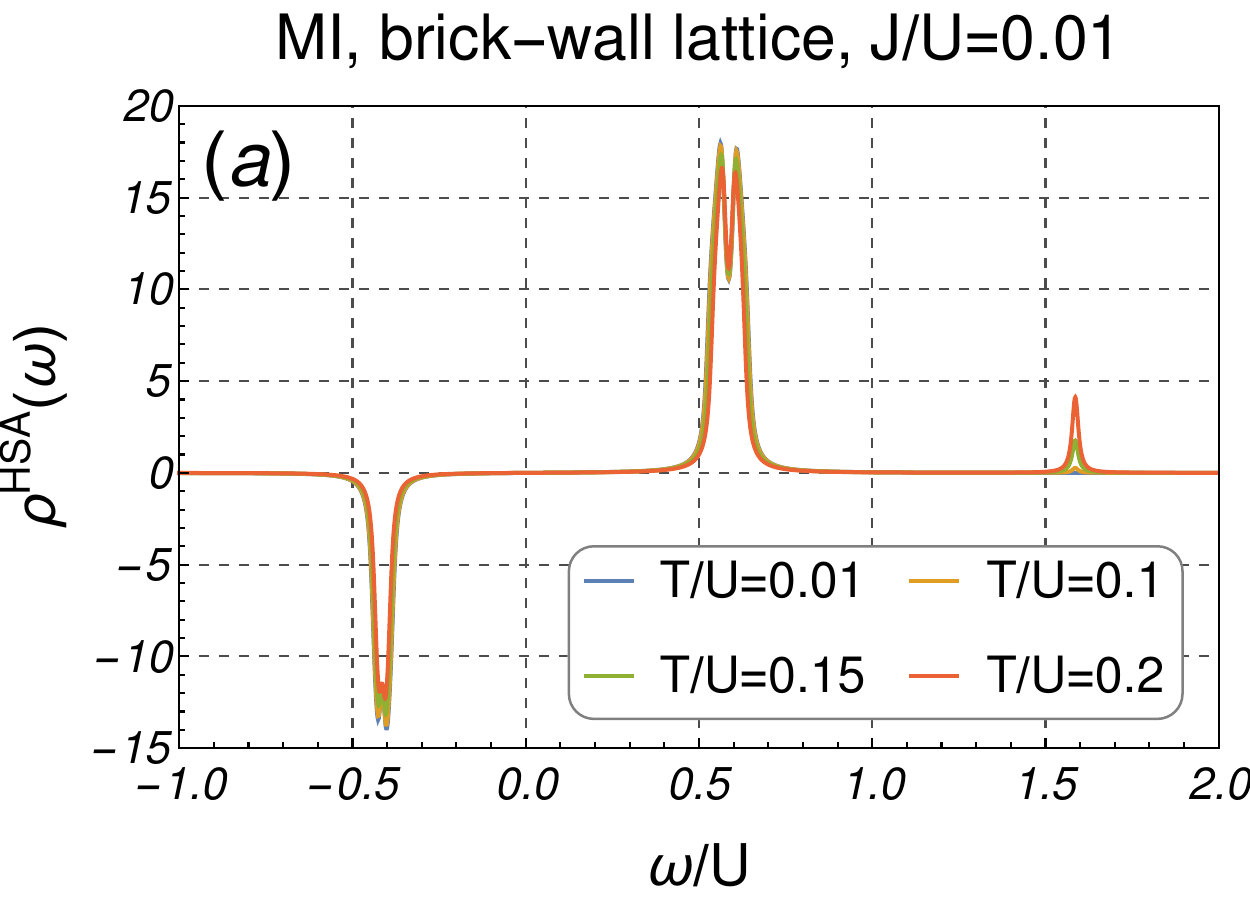}
\includegraphics[scale=0.45]{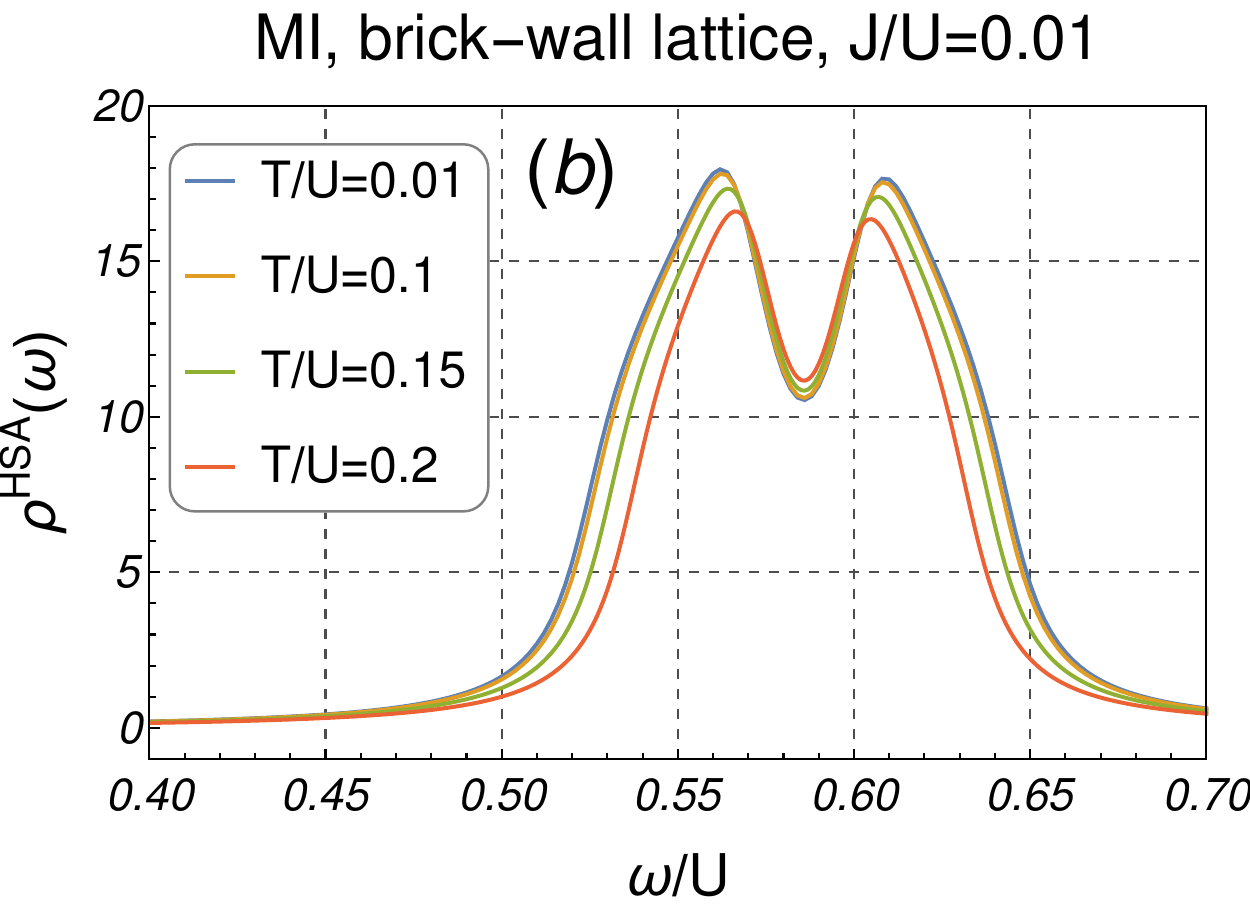}
\includegraphics[scale=0.45]{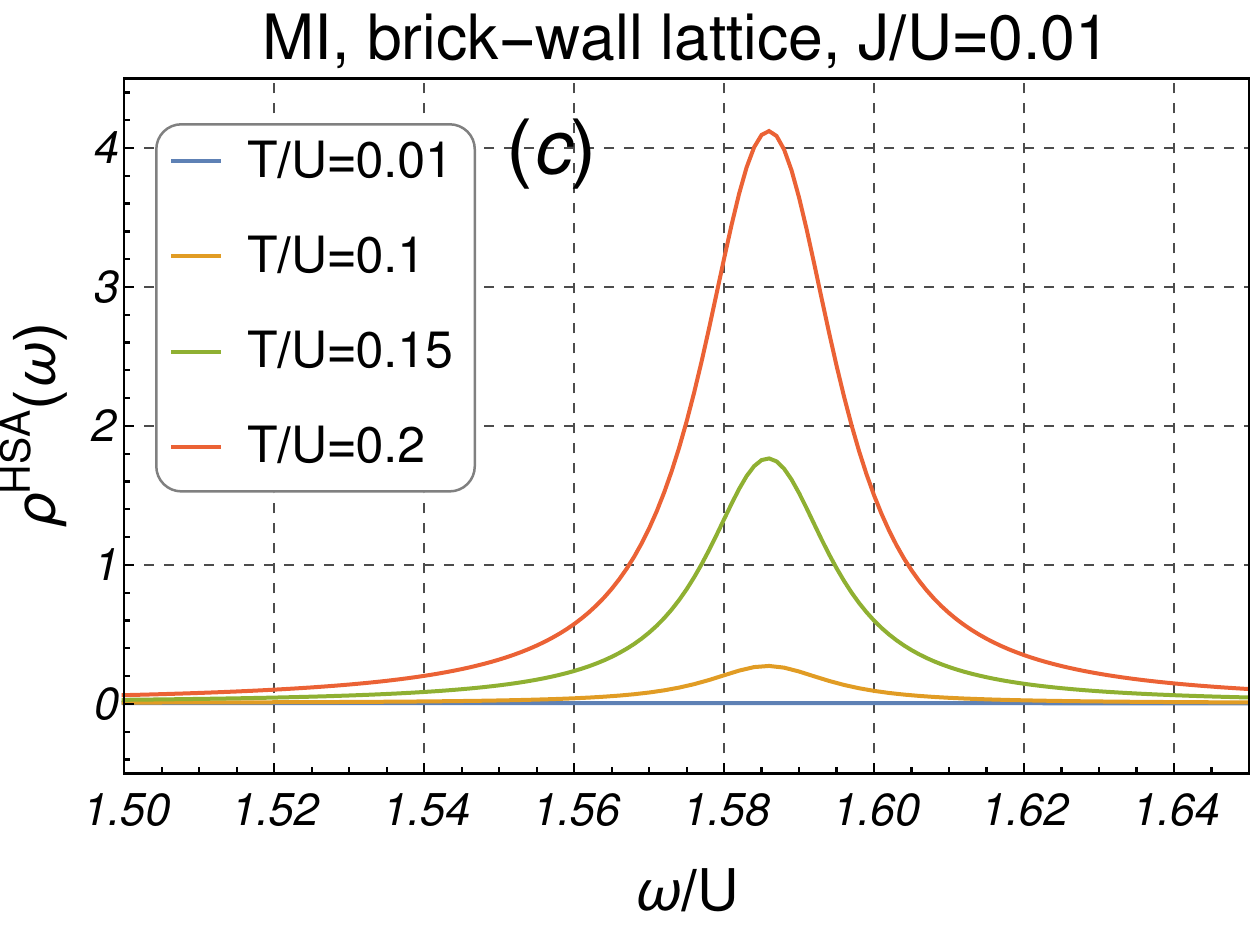}

\includegraphics[scale=0.45]{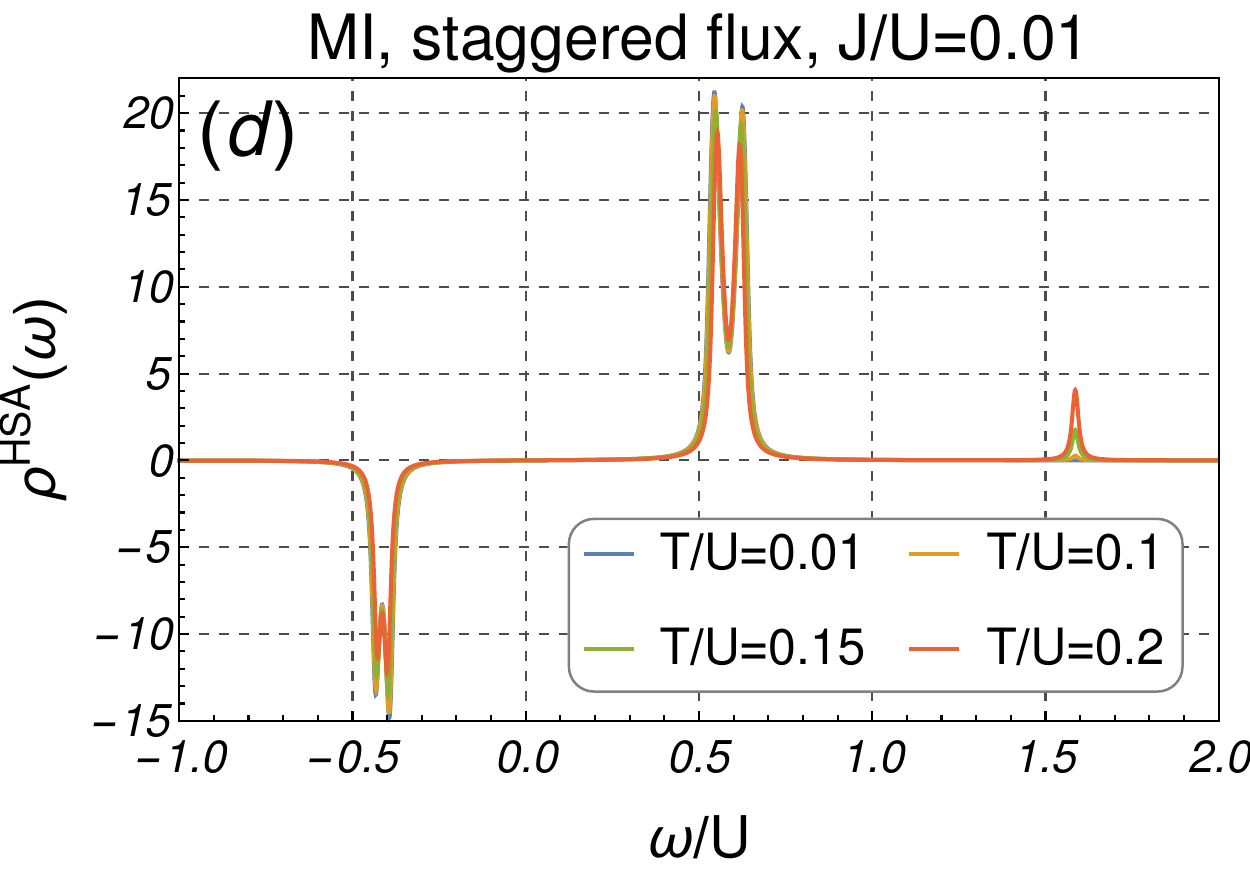}
\includegraphics[scale=0.45]{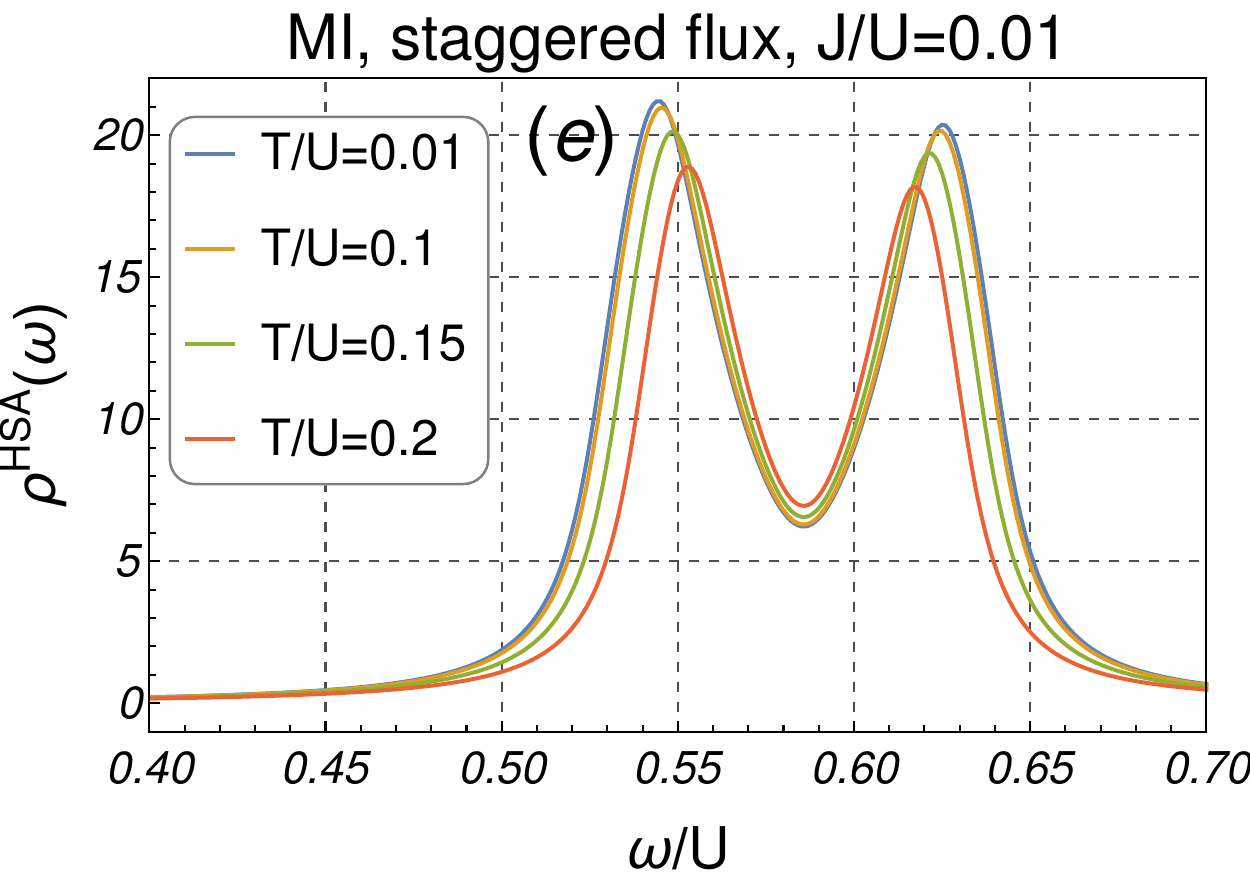}
\includegraphics[scale=0.45]{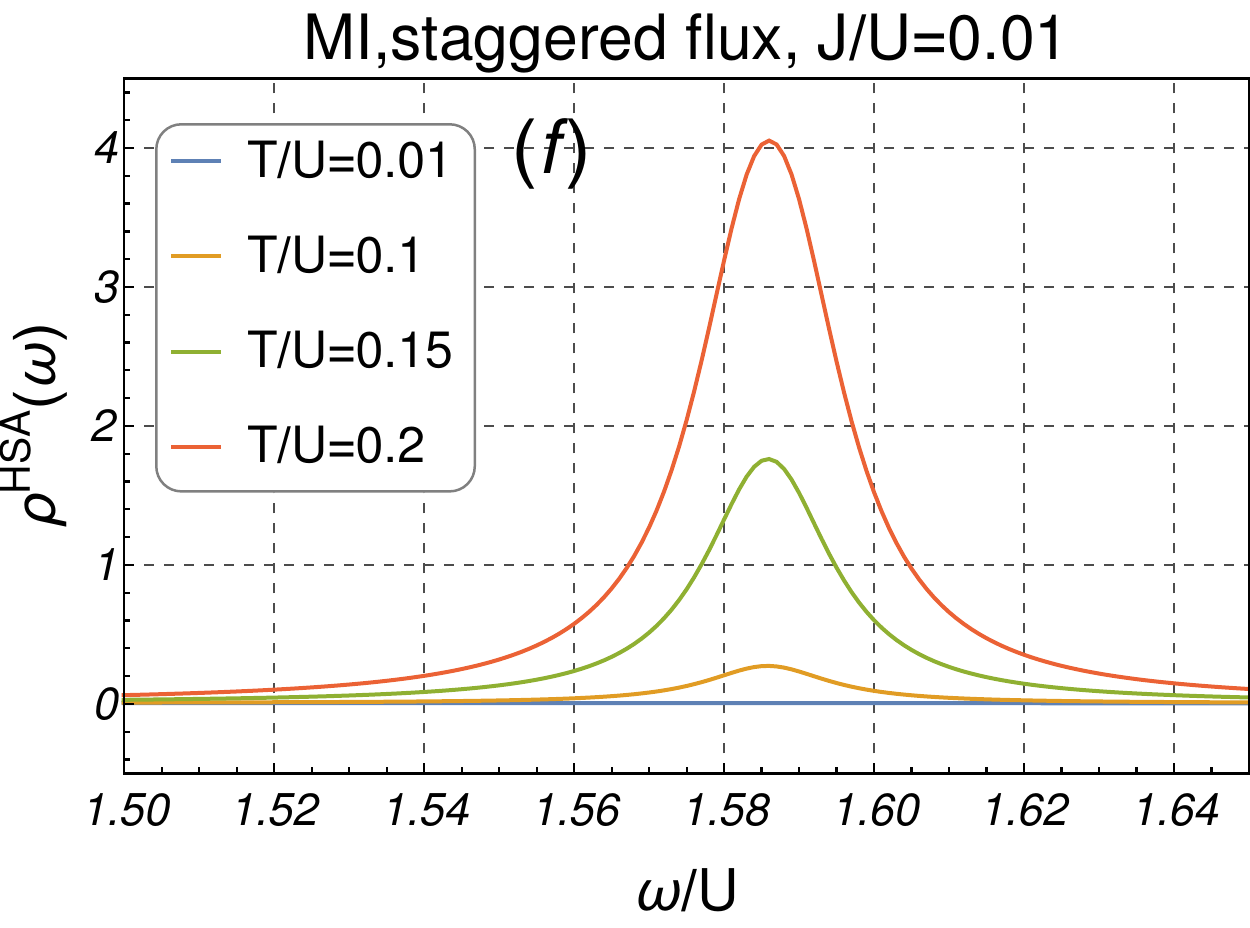}

\caption{(color online) Finite temperature density of states in bosonic MI
phase; (a-c) correspond to the brick wall lattice and (d-f) correspond
to the staggered flux lattice. The plots are drawn for the different
temperature values: $T/U=0.01$ - blue, $T/U=0.1$ - orange, $T/U=0.15$
- green, $T/U=0.2$ - red line. Moreover, the deep MI regime was chosen,
i.e. $J/U=0.01$. \label{fig: finite- TEMP}}
\end{figure*}

\subsection{Quasi-particle density of states at finite temperatures}

To describe quasi-particle DOS at finite temperatures, we use the
higher state approximation (HSA) for the local Green function $G_{0}(i\omega_{m})=G_{0}^{HSA}(i\omega_{m})$
(see Appendix \ref{sub:Local-Green-function} and Eq. (\ref{eq:HSAg0}))
\cite{PhysRevA.94.043612}. This form of $G_{0}$ takes into account
thermal fluctuations which are observed in the periodically modulated
lattice experiments \cite{PhysRevA.94.043612,PhysRevLett.92.130403}.

Applying similar a procedure as in Sec. \ref{sub:Single-particle-density T=00003D0}
but with the local Green function $G_{0}^{HSA}(i\omega_{m})$, we
get\begin{widetext}
\begin{eqnarray}
\rho_{MI}^{HSA}\left(\omega\right) & = & -\frac{1}{\pi N}\sum_{\lambda=1}^{N}\textrm{Im}\left[\frac{L\left(E_{h}(\varepsilon_{\lambda})\right)}{\left(\omega-E_{h}(\varepsilon_{\lambda})+i\eta\right)\left(E_{h}(\varepsilon_{\lambda})-E_{p}(\varepsilon_{\lambda})\right)\left(E_{h}(\varepsilon_{\lambda})-E_{t}(\varepsilon_{\lambda})\right)}\right.\nonumber \\
 &  & +\frac{L\left(E_{p}(\varepsilon_{\lambda})\right)}{\left(\omega-E_{p}(\varepsilon_{\lambda})+i\eta\right)\left(E_{p}(\varepsilon_{\lambda})-E_{h}(\varepsilon_{\lambda})\right)\left(E_{p}(\varepsilon_{\lambda})-E_{t}(\varepsilon_{\lambda})\right)}\nonumber \\
 &  & \left.+\frac{L\left(E_{t}(\varepsilon_{\lambda})\right)}{\left(\omega-E_{t}(\varepsilon_{\lambda})+i\eta\right)\left(E_{t}(\varepsilon_{\lambda})-E_{h}(\varepsilon_{\lambda})\right)\left(E_{t}(\varepsilon_{\lambda})-E_{p}(\varepsilon_{\lambda})\right)}\right]\,,\label{eq: main-result HSA}
\end{eqnarray}
\end{widetext}where
\begin{equation}
E_{p/h}(\varepsilon_{\lambda})=A(\varepsilon_{\lambda})-B(\varepsilon_{\lambda})\sin\left(\frac{\pi}{6}\mp\frac{1}{3}\arccos C(\varepsilon_{\lambda})\right),\label{eq:E2}
\end{equation}

\begin{equation}
E_{t}(\varepsilon_{\lambda})=A(\varepsilon_{\lambda})+B(\varepsilon_{\lambda})\cos\left(\frac{1}{3}\arccos C(\varepsilon_{\lambda})\right),\label{eq:E1}
\end{equation}
and the definitions of $A(\varepsilon_{\lambda})$, $B(\varepsilon_{\lambda})$,
$C(\varepsilon_{\lambda})$, $L\left(x\right)$ are given in Appendix
\ref{sub:HSA-coefficients}. The above calculations extend the results
obtained previously for the cubic lattice with periodic boundary conditions
from Ref. \cite{PhysRevA.94.043612}.

In the following, we exploit the above method to study the finite-size
lattices currently accesible in the optical lattice experiments, i.e.
we focus on the square and brick-wall geometry and the lattice with
a staggered magnetic field \cite{2012Natur.483..302T,PhysRevLett.110.165304,2010PhRvA..81b3404L,2008PhRvL.100m0402L}.

\section{Results}

\subsection{Phase diagram at zero temperature limit \label{sub:Phase-diagram-at}}

Before we follow the considerations which involve the analysis of
quasi-particle spectra in the MI phase, firstly we have to assess
the range of Hamiltonian parameters for which the MI phase is a ground
state of BHM. This will be done by investigating the phase boundary
between MI and SF phase at zero temperatures by using Eq. (\ref{eq:phase-boundary}).
As we mentioned before we are interested in the different lattice/gauge
geometries, like square and brick-wall type, and the lattice with
staggered magnetic flux. Moreover, the particle hopping on these lattices
will be affected by the Landau gauge which result in the self-similar
structure encoded in the particle excitations. It can be achieved
by proper incorporation of the hopping matrix $\mathbf{J}$ into the
effective action $S^{eff}$ from Eq. (\ref{eq:Seff3}) and by introduction
of Peierls substitution to it \cite{PhysRevLett.108.225303,2014PhRvA..89b3631S,Kopec:2002vy},
i.e.
\begin{equation}
J_{ij}\rightarrow J_{ij}e^{i2\pi\int_{j}^{i}\mathbf{A}_{L}\cdot d\mathbf{l}}
\end{equation}
where the Landau gauge is defined as $\mathbf{A}_{L}=B(0,x,0)$ (with
flux per plaquette $p/q$ defined as $p/q=\eta Ba^{2}$ where $a$
is the lattice spacing and $\eta=1$ ($\eta=2$) for the square (brick-wall)
lattice). Moreover, the staggered flux lattice represented by the
gauge $\mbox{\ensuremath{\mathbf{A}}}_{S}$ \cite{2008PhRvL.100m0402L,2010PhRvA..81b3404L},
together with Landau gauge $\mathbf{A}_{L}$, are introduced to the
system in superposition, i.e. $\mathbf{A}_{L}+\mbox{\ensuremath{\mathbf{A}}}_{S}$.
Formally, the hopping around the plaquette with vertices 1, 2, 3,
4 for $\mbox{\ensuremath{\mathbf{A}}}_{S}$ is defined as $-Je^{i\phi/4}\left(a_{2}^{\dagger}a_{1}+a_{3}^{\dagger}a_{2}+a_{4}^{\dagger}a_{3}+a_{1}^{\dagger}a_{4}\right)$
which yields $\phi$ flux, see Fig. \ref{fig: gauges} c. Moreover,
for clarity, we plot the relevant lattices and gauges in Figs. \ref{fig: gauges}
a and b.

The phase diagrams for the finite size lattices (30x30 sites) without
uniform synthetic magnetic field ($p/q=0$) are presented in Fig.
\ref{fig: pb-lobes}. On the mean-field level, the phase boundaries
for the square and staggered flux lattices, have been earlier analyzed
in Refs. \cite{vanOosten:2001kp,Sajna2015,2010PhRvA..81b3404L} (phase
boundary for the brick-wall lattice is the same as in the honeycomb
case \cite{PhysRevA.92.043605,PhysRevA.87.043619}). As we see, the
brick-wall and staggered flux lattices favor the MI phase in which
the atoms are likely to be localized at the lattice sites. This effect
could be simply accounted for the fact that both types of lattices
break translational symmetry and the elementary unit cell contains
two lattice sites.

Next, we focus on a more interesting situation, i.e. when the uniform
synthetic magnetic field is applied in the whole relevant range of
$p/q$. Moreover in our analysis, we also consider the finite size
effects the lattice pattern which will be discussed in the following.

Fig. \ref{fig:phase boundary at tip} gives a plot of the phase boundary
at the tip of the first lobe versus flux per plaquette. Firstly, in
the absence of uniform magnetic field, we compare the phase boundaries
for the standard square lattice ($\phi=0$) (a), staggered flux lattice
($\phi=\pi$) (b), and brick-wall lattice (d). For the three lattices,
as expected, the critical lines show different behavior. Especially
we see that the critical lines for the square (Fig. \ref{fig:phase boundary at tip}
a) and brick-wall (Fig. \ref{fig:phase boundary at tip} d) lattices
are shifted with respect to each other by $p/q=1/2$ value (in the
next paragraph we provide more detailed discussion of this). Moreover,
it is also interesting to notice that the finite size effects are
quite small for the lattices of 30x30 size for which the $p/q$ phase
boundary dependence is less blurred out (compare this with 6x6 or
10x10 lattice sizes in Fig. \ref{fig:phase boundary at tip} and with
periodic boundary condition calculations in Ref. \cite{PhysRevB.75.045133}
in which standard square lattice was considered). Therefore, our calculations
indicate that non-monotonous critical line behavior is quite well
resolved on the 30x30 lattice size which is useful information for
the future experimental setups (30x30 lattice size systems are currently
realizable, e.g. see \cite{Bakr2011,Bakr:2010wn}).

As we mentioned above, we provide here a more detailed discussion
of the critical line versus $p/q$ dependence for different staggered
flux amplitudes which are parametrized by $\phi$ (see Figs. \ref{fig:phase boundary at tip}
a-c). In Figs. \ref{fig:phase boundary at tip} a and c, the two limits
of $\phi$, i.e. $\phi=0$ and $\phi=\pi$ correspond to the standard
square lattice and the staggered flux lattice with maximal value of
flux per plaquette, respectively. As we pointed out above both critical
lines are shifted to each other by $p/q=1/2$. It can be argued on
the background of the tight binding dispersions whose analytical forms
for the lattice with periodic boundary conditions are known. Namely,
for the case $\phi=0$ (Fig. \ref{fig:phase boundary at tip} a),
uniform magnetic field linearly increase from $p/q=0$ up to maximal
value at $p/q=1/2$ and linearly decrease from $p/q=1/2$ up to $p/q=1$.
This results in the symmetrical form of the phase boundary around
the $p/q=1/2$ point. In particular at $\phi=0$ for $p/q=0$ and
$p/q=1$ lattice dispersion is 
\begin{equation}
\varepsilon\left(k_{x},\, k_{y},\,\phi=0,\, p/q=0\,\textrm{or\,}1\right)=2\left(\cos k_{x}+\cos k_{y}\right)\label{eq: dispersion square lattice}
\end{equation}
and at point $p/q=1/2$ the dispersion shows two branches \cite{2014PhRvA..89b3631S}
\begin{equation}
\varepsilon\left(k_{x},\, k_{y},\,\phi=0,\, p/q=1/2\right)=\pm2\sqrt{\cos^{2}k_{x}+\cos^{2}k_{y}}.\label{eq: dispersion f=00003D1/2}
\end{equation}
where $k_{x}$ and $k_{y}$ are wave vectors. In contrast, for $\phi=\pi$
(Fig. \ref{fig:phase boundary at tip} b) the sweeping range of $p/q$
parameter is the same but at $p/q=0$ and $p/q=1$ lattice dispersion
is \cite{2008PhRvL.100m0402L,2010PhRvA..81b3404L} 
\begin{eqnarray}
 &  & \varepsilon\left(k_{x},\, k_{y},\,\phi=\pi,\, p/q=0\right)\label{eq:}\\
 &  & =\pm2\sqrt{\cos^{2}\left(\frac{k_{x}+k_{y}}{2}\right)+\cos^{2}\left(\frac{k_{x}-k_{y}}{2}\right)},\label{eq:dispersion phi=00003DPi}
\end{eqnarray}
which has the same lowest energy behavior as for $\phi=0$ at $p/q=1/2$
point (see Eq. (\ref{eq: dispersion f=00003D1/2})). Moreover, the
point $p/q=1/2$ for $\phi=\pi$ and its lowest energy behavior is
the same as in $\phi=0$ for $p/q=0$ and $p/q=1$ (Eq. \ref{eq: dispersion square lattice}).
Therefore, we see that these points, and also the rest of the critical
lines in Fig. \ref{fig:phase boundary at tip} a) and b) correspond
to each other and are shifted by $p/q=1/2$.

However the correspondence between the phase boundary shapes obtained
for the standard square lattice ($\phi=0$) and the staggered flux
lattice ($\phi=\pi$) for different $p/q$ parameters disappears when
the phase boundary is analyzed for the intermediate values of $\phi$,
i.e. for $\phi=\pi/4$, $\phi=\pi/2$, $\phi=3\pi/4$ (see, Fig. \ref{fig:phase boundary at tip}
c). Then the critical line changes nonlinearly between the values
$\phi=0$ and $\phi=\pi$ and we cannot provide any intuitive interpretation
for this situation. 

Moreover, we numerically discover that at the points $p/q=1/4$ and
$p/q=3/4$ all critical lines for different $\phi$ almost intersect
each other (see arrows in Fig. \ref{fig:phase boundary at tip} c). This intersection area becomes smaller with increasing
lattice size. It is confirmed by the results presented in Fig. \ref{fig: pb-special-point-1/4}
which show that with increasing lattice size (here up to 200x200 lattice
sites) the dependence of critical line at $p/q=1/4$ for different
$\phi$ value is almost constant (see the scale on the vertical axis
in Fig. \ref{fig: pb-special-point-1/4}). This simply shows that
the tip of the first lobe in the phase diagram is intact when staggered
flux of different amplitudes are applied to the system and this only
happens when the system is subjected to a uniform magnetic field with
$p/q=1/4$ or $p/q=3/4$ strength.

Summarizing this subsection, we have shown that a combination of the
Hofstadter butterfly spectrum with a staggered flux background gives
nontrivial MI-SF critical behavior. In particular, this behavior is
highly nonlinear and exhibits robustness in some range of Hamiltonian's
parameters. For example, the latter effect could be of interest in
quality testing of gauge field within the critical region in the experimental
protocol. In further discussion we focus on the bosonic MI phase and
its spectral properties, which is the main subject of concern in our
study.

\subsection{Hofstadter butterfly in the MI phase for different lattice geometry}

Here we analyze the dependence of the MI spectrum in the whole range
of the uniform synthetic magnetic field strength within the square
and brick-wall lattices. The brick-wall lattice is especially interesting
because the relativistic dispersion appears in its single-particle
spectrum, which will be important in further discussion. For the two
lattices chosen, we set the number of sites to be $30\times30$\textcolor{black}{{}
and as could be seen in }Fig. \ref{fig: butterfly FREE},\textcolor{black}{{}
the obtained pattern for the free bosonic case ($U=0$) resembles
the previous works \cite{1976PhRvB..14.2239H,PhysRevA.91.063628,PhysRevA.76.055601}
(the calculations were made by using Eq. (\ref{eq: FREE DOS})). To
realize one of the main aim of the study, we used these spectra to
evaluate the strongly correlated density of states in the MI phase}
with unit average density ($n_{0}=1$). It was achieved by using Eq.
(\ref{eq: main-result}) and the density plots of these spectra in
terms of fluxes per plaquette ($p/q$) versus frequencies $\omega/U$
are depicted in Fig. \ref{fig:butterfly}. 

As follows from this figure(Fig. \ref{fig:butterfly}) the free particle
picture is changed when the interactions between bosons are taken
into account $U/J\gg1$. It is manifested by opening a characteristic
MI energy gap around $\omega/U=0$. The gap divides the quasi-particle
spectrum into two branches of the hole and particle like character.
Moreover, we see that each of these branches\textcolor{red}{{} }also
show self-similar behavior in which the Hofstadter butterfly structure
is restored. In particular in the square lattice (Figs. \ref{fig:butterfly}
a, b, c, d) the hole (for $\omega/U<0$) and particle (for $\omega/U>0$)
branches acquire the self-similar structure which corresponds to the
free particle case from Fig. \ref{fig:butterfly} a (the adequate
correspondence in the brick-wall lattice is between the spectra in
Figs. \ref{fig:butterfly} e, f, g, h and Fig. \ref{fig:butterfly}
b). However, deep in the Mott phase in which hopping kinetic energy
is weak ($J\ll U$), the Hofstadter butterfly like behavior is not
well resolved due to a finite size effect and spectral broadening
of the parameter $\eta$ (see Figs. \ref{fig:butterfly} a, b, e,
f). Therefore, future experimental setups which investigate such a
strongly correlated system could partially overcome this problem by
adjusting experimental parameters close to the SF-MI phase boundary,
as in Figs. \ref{fig:butterfly} d and h, for which quasi-particle
bandwidth broadens and $J/U$ is close to the critical value (see,
Fig. \ref{fig: pb-special-point-1/4} a and d).

Finally, we can conclude that the lattice geometry and strong interactions
have a significant impact on the self-similar structure of the quasi-particle
MI spectra. Further, we show that the tunability of the self-similar
spectrum of the strongly interacting bosons can be widely enhanced
going beyond the geometrical modification of the lattice, i.e. by
using staggered flux lattice.

\subsection{Hofstadter butterfly in MI phase with staggered magnetic field background
\label{sub:Hofstadter-butterfly-in staggered}}

In this subsection we focus on the staggered flux lattice with checkerboard
symmetry \cite{2008PhRvL.100m0402L,2010PhRvA..81b3404L}. We use Eqs.
(\ref{eq: main-result}-\ref{eq: FREE DOS}) to obtain DOS for whole
range of the uniform magnetic field amplitude (see Fig. \ref{fig:butterfly-1}).
The non-interacting and interacting self-similar pattern are plotted
in Figs. \ref{fig:butterfly-1} a-d and e-h, respectively. In this
figures, we show the Hosftadter butterflies for different flux per
plaquette $\phi$ with chosen $J/U$. In particular, we see that different
kind of fractal like pattern emerges when amplitude of $\phi$ varies.
This behavior is nonlinear (similar conclusion for phase boundary
analysis was made in Sec. \ref{sub:Phase-diagram-at}). However, one
can see that self similar structure for $\phi=0$ (Fig. \ref{fig: butterfly FREE}
a and \ref{fig:butterfly} a-d) and $\phi=\pi$ (Fig. \ref{fig:butterfly-1}
d and h) cases are shifted each other by the $p/q=1/2$ which agrees
with our earlier observation in Sec. \ref{sub:Phase-diagram-at}. 

Moreover, this is important to notice here that although the quasi-particle
spectra in the MI phase for staggered flux lattice and brick-wall
lattice show qualitatively similar DOS (see Fig. \ref{fig: DOS exact - square and brick-wall}
b) and Fig. \ref{fig: DOS exact - staggered} a and b), they give
completely different Hofstadter butterfly patterns of DOS when a uniform
magnetic field is turned on. This simple argument shows that mechanism
of self-similarity behavior is much more complex and DOS picture is
not an adequate tool to build up an intuition about this peculiar
phenomenon. Additionally, in Figs. \ref{fig: DOS exact - square and brick-wall}
and \ref{fig: DOS exact - staggered} we compare some chosen quasiparticle
DOS of the finite size lattice systems with exact DOS obtained for
the periodic boundary conditions (see also Appendix \ref{sub:Exact-density-of})
with $p/q=0$. This shows that 30x30 lattice size quite well reproduces
the shape of DOS also in the region around Dirac points i.e. at which
DOS value is highly suppressed, Fig. \ref{fig: DOS exact - square and brick-wall}
b and Fig. \ref{fig: DOS exact - staggered} a and b).

As follows from the above results, the tunability of the lattice in
non-geometrical way, e.g. through a gauge field, can be also interesting
in studying of nontrivial phenomena. Such a staggered flux lattice
is a good example although it has a relatively simple Abelian structure. 

To better embed the above results in the context of real experimental
system, we focus further on the problem of thermal fluctuations. Namely,
such fluctuations are difficult to control in the experimental protocol,
especially in the strongly interacting limit \cite{Trotzky:2010ti,Zhou2011}.
Therefore, in the following, we provide an analysis of these effects
within HSA approach.

\subsection{Quasiparticle excitations at finite temperatures }

Optical lattice experiments are not conducted at the strictly zero
temperature regime \cite{Trotzky:2010ti,Bakr:2010wn,Sherson:2010wz},
therefore it is important to consider the effects of thermal fluctuations.
I is particularly interesting to investigate a region around Dirac
points in Fig. \ref{fig: DOS exact - square and brick-wall} b and
Fig. \ref{fig: DOS exact - staggered} a and b. Here we focus on the
whole relevant range of temperatures in the deep MI phase, i.e. we
investigate temperatures up to $T/U=0.2$ which is melting point of
MI phase \cite{Gerbier:2007to,Polak-temp}. It should be added that
the MI phase at a finite temperature does not exist, however it is
justified to discuss the MI phase below $T/U=0.2$ because MI properties
can be observed up to this temperature \cite{Gerbier:2007to,Polak-temp}.

To properly catch the finite temperature regime in the MI phase, higher
order energy states should be taken into account. because these states
get occupied due to thermal fluctuations \cite{PhysRevA.94.043612}.
In the context of the lattices with relativistic dispersion, i.e.
brick-wall and staggered flux lattices, we investigate this phenomena
with the HSA and calculate DOS by using Eq. (\ref{eq: main-result HSA}).
The results are depicted in a-c for brick-wall and d-f for staggered
flux lattice. In particular, the first two peaks in Figs. \ref{fig: finite- TEMP}
a and d correspond to the holon and doublon excitations (from the
left). The third peak in Figs. \ref{fig: finite- TEMP} a and d is
a fingerprint of the triplon defects over the MI ground state. From
these diagrams, two main features follow: 1) depletion of the holon
and doublon excitations at the expense of triplons ones at high temperatures
(see also Figs. \ref{fig: finite- TEMP} b, c and e, f), 2) the vicinity
of the Dirac points are highly robust against the thermal fluctuations
(see, Figs. \ref{fig: finite- TEMP} b and e). The latter observation
gives the important information for the future experimental setups
which shows that the interesting Dirac like physics could still be
accessed even at relatively high temperatures ($T/U<0.2$).

\section{Summary}

We have applied the strong coupling expansion method to the BHM in
the finite size lattices, which allowed us to study the energy spectrum
in the arbitrary gauge fields and at finite temperatures. As an example
we focused on the lattices with relativistic dispersions i.e. on the
brick-wall and staggered flux ones. In particular, we have shown that
strong on-site interaction of bosons does not destroy self-similar
like structures in the quasi-particle spectrum, known as a Hofstadter
butterfly but modifies them significantly. We have also noticed that
- quasi-particle density of states for both types of lattices studied
are qualitatively similar, however they give completely different
self-similar pattern, when staggered flux amplitude is tuned. This
analysis was performed for a gapped phase of the Bose Hubbard model
(i.e. in MI phase). Additionally, we have presented that such a remarkable
fractal patterns can be only efficiently studied in the vicinity of
the phase boundary, because of widening of the quasi-particle and
hole energy bands for which magnetic flux dependence on energy scales
is better resolved. Moreover, we have numerically shown that the phase
boundary is intact over all range of staggered flux amplitudes within
the uniform magnetic field at $p/q=1/4$ and $p/q=3/4$. It indicates
that simple superposition of two different synthetic magnetic fields
can be a generator of non-trivial phenomena in the optical lattice
systems.

In this work, we have also focused on the quasi-particle excitations
at finite temperatures and investigated how they are modified by thermal
defects. This investigation is especially important for the experimental
realization of the gauge fields in which e.g. Dirac like physics emerges
(see Ref. \cite{Goldman2014} and literature therein). In particular,
we have shown that the vicinities of the Dirac points in DOS are highly
robust against thermal fluctuations and can be efficiently studied
in the experimental setups even at relatively high temperatures (up
to MI melting point $T/U=0.2$ \cite{Gerbier:2007to}).

Moreover, by including finite size effects, we have simulated the
lattice sizes which are currently accesible in experimental protocols
\cite{Bakr:2010wn,Bakr2011} and we have shown that this sizes are
sufficient to observe orbital magnetic field phenomena in the BHM.
\begin{acknowledgments}
We are grateful to P. Ro\.{z}ek for valuable discussions. This work
was supported by the National Science Centre, Poland, project no.
2014/15/N/ST2/03459 (A. S. S.).
\end{acknowledgments}

\section{Appendix}

\subsection{Local Green function \label{sub:Local-Green-function}}

General form of the local Green function from Eq. (\ref{eq:local-green})
could be rewritten in the Matsubara frequencies as a

\begin{equation}
\frac{1}{\hbar}G_{0}\left(i\omega_{m}\right)=-\sum_{n_{0}=0}^{\infty}\frac{\left(n_{0}+1\right)\left(f_{n_{0}+1}-f_{n_{0}}\right)}{i\hbar\omega_{m}-\left(E_{n_{0}+1}-E_{n_{0}}\right)},\label{eq:G0}
\end{equation}
\begin{equation}
f_{k}=\frac{e^{-\beta E_{k}}}{\sum_{n_{0}=0}^{\infty}e^{-\beta E_{n_{0}}}}\,,\label{eq:fn-definicja}
\end{equation}
where $E_{n_{0}}=-\mu n_{0}+Un_{0}\left(n_{0}-1\right)/2$ and $n_{0}$
is an integer value denoting local occupation number of bosons per
site. 

In this paper, we consider three (TSA) \cite{2008PhRvB..77w5120M,Ohashi:2006ui,Freericks-strong-coupling,PhysRevA.73.033621,2005PhRvA..71c3629S,vanOosten:2001kp,PhysRevA.68.043623,Sajna2015,2014PhRvA..89b3631S}
and higher state (HSA) \cite{PhysRevA.94.043612} approximations which
correspond to the truncation of the sum in Eq. (\ref{eq:G0}) to the
three (TSA) or more (HSA) indices around some chosen value of $n_{0}$,
respectively (it turns out that this is sufficient approximation to
describe MI state at low temperatures \cite{PhysRevA.80.033612,Ohliger2013}).
However, at finite temperatures both of approximations gives the same
zero temperature limit of $G_{0}\left(i\omega_{m}\right)$, i.e.
\begin{eqnarray}
 & \frac{1}{\hbar}G_{0}^{TSA}\left(i\omega_{n}\right)= & \frac{n_{0}+1}{i\hbar\omega_{m}-\left(E_{n_{0}+1}-E_{n_{0}}\right)}\nonumber \\
 &  & -\frac{n_{0}}{i\hbar\omega_{m}-\left(E_{n_{0}}-E_{n_{0}-1}\right)}\,,\label{eq:TSAg0}
\end{eqnarray}
where $f_{n_{0}}=1$ and $f_{n_{0}-1}=f_{n_{0}+1}=0$. The form of
$G_{0}^{TSA}\left(i\omega_{n}\right)$ is known as TSA because it
take into account three possible bosonic occupation numbers denoted
by $n_{0}-1$, $n_{0}$, $n_{0}+1$. At finite temperature the situation
is more involved. It turns out that higher order approximations are
needed to correctly describe thermal fluctuations effects observed
e.g. in periodically driven optical lattice systems \cite{PhysRevA.94.043612,PhysRevLett.92.130403}.
Then, when the system is e.g. in MI phase with unit density ($n_{0}=1$),
it is enough to choose $n_{0}=0,\,1,\,2,\,3$ which result in the
following form of the local Green function within HSA method
\begin{eqnarray}
 &  & \frac{1}{\hbar}G_{0}^{HSA}\left(i\omega_{m}\right)=-\frac{f_{1}-f_{0}}{i\hbar\omega_{m}-\left(E_{1}-E_{0}\right)}\nonumber \\
 &  & -\frac{2\left(f_{2}-f_{1}\right)}{i\hbar\omega_{m}-\left(E_{2}-E_{1}\right)}-\frac{3\left(f_{3}-f_{2}\right)}{i\hbar\omega_{m}-\left(E_{3}-E_{2}\right)}\,.\label{eq:HSAg0}
\end{eqnarray}

\subsection{HSA coefficients of density of states \label{sub:HSA-coefficients}}

Coefficients in Eqs. (\ref{eq: main-result HSA}-\ref{eq:E1}) have
the following forms

\begin{widetext}
\begin{equation}
A\left(\varepsilon_{\lambda}\right)=\frac{1}{3}\left(\varepsilon_{\lambda}+3U-3\mu-4\varepsilon_{\lambda}f_{3}\right),
\end{equation}
\begin{equation}
B(\varepsilon_{\lambda})=\frac{2}{3}\sqrt{3U^{2}+\varepsilon_{\lambda}\left(1-4f_{3}\right){}^{2}+3\varepsilon_{\lambda}U\left(-1+2f_{1}+4f_{2}-2f_{3}\right)},
\end{equation}
\begin{equation}
C(\varepsilon_{\lambda})=-\frac{\varepsilon_{\lambda}\left(9\varepsilon_{\lambda}U\left(-1+2f_{1}+4f_{2}-2f_{3}\right)\left(-1+4f_{3}\right)+2\varepsilon_{\lambda}^{2}\left(-1+4f_{3}\right){}^{3}+9U^{2}\left(-1+6f_{1}-6f_{2}+4f_{3}\right)\right)}{2\left(3U^{2}+\varepsilon_{\lambda}^{2}\left(1-4f_{3}\right){}^{2}+3\varepsilon_{\lambda}^{2}U\left(-1+2f_{1}+4f_{2}-2f_{3}\right)\right){}^{3/2}},
\end{equation}
\begin{equation}
L\left(x\right)=(U-\mu-x)(2U-\mu-x)f_{0}-(2U-\mu-x)(U+\mu+x)f_{1}+(\mu+x)\left((U+\mu+x)f_{2}+3(U-\mu-x)f_{3}\right)\,,\label{eq:G_L}
\end{equation}
\end{widetext}where $f_{m}$ are defined in Eq. (\ref{eq:fn-definicja}).

\subsection{Exact density of states \label{sub:Exact-density-of}}

For depicting of the exact quasi-particle DOS in Fig. \ref{fig: DOS exact - square and brick-wall}
and \ref{fig: DOS exact - staggered} we used the following lattice
DOS for free bosons (we set units $J=1$):

- DOS for square lattice with dispersion $2\left(\cos k_{x}+\cos k_{y}\right)$
takes the form
\begin{equation}
\rho_{sq}(\omega)=\frac{1}{\pi^{2}2}\mathcal{K}\left(\sqrt{1-\left(\frac{\omega}{4}\right)^{2}}\right)\,,
\end{equation}

- DOS for brick-walll lattice with dispersion $2\sqrt{\cos k_{x}\cos k_{y}+\cos^{2}k_{x}+\frac{1}{4}}$
\cite{PhysRevLett.110.165304} takes the form
\begin{equation}
\rho_{bw}(\omega)=\frac{1}{4\pi^{2}}\int_{-1}^{1}\frac{\left|\omega\right|\theta\left(3^{2}-\omega^{2}\right)du}{\left|u\right|\sqrt{1-u^{2}}\sqrt{1-\left(\frac{\left(\frac{\omega}{2}\right)^{2}-u^{2}-\frac{1}{4}}{u}\right)^{2}}}\,,
\end{equation}

- DOS for staggered flux lattice with dispersion $2\sqrt{2\cos\left(\frac{\phi}{2}\right)\cos k_{+}\cos k_{-}+\cos^{2}k_{+}+\cos^{2}k_{-}}$
\cite{2008PhRvL.100m0402L,2010PhRvA..81b3404L} where $k_{+}=\left(k_{x}+k_{y}\right)/2$,
$k_{-}=\left(k_{x}-k_{y}\right)/2$ takes the 
\begin{eqnarray}
\rho_{st}(\omega) & = & \frac{1}{4\pi^{2}}\int_{-1}^{1}\frac{1}{\sqrt{1-u^{2}}\left|u+\cos\left(\frac{\text{\ensuremath{\phi}}}{2}\right)\right|}\nonumber \\
 &  & \times\frac{\left|\omega\right|du}{\sqrt{1-\left(\frac{\left(\frac{\omega}{2}\right)^{2}-u\cos\left(\frac{\text{\ensuremath{\phi}}}{2}\right)-1}{\cos\left(\frac{\text{\ensuremath{\phi}}}{2}\right)+u}\right)^{2}}}\,.
\end{eqnarray}

\bibliographystyle{apsrev}
\bibliography{library}

\end{document}